\documentclass[aps,pra,twocolumn,showpacs,superscriptaddress,floatfix,nofootinbib]{revtex4}
\usepackage{graphicx}
\usepackage{amsmath}
\usepackage{epsfig}
\usepackage{helvet}
\usepackage{amssymb}

\begin{document}

\title[Entanglement renormalization in free bosonic systems]{Entanglement renormalization in free bosonic systems: real-space versus momentum-space renormalization group transforms}
\author{G. Evenbly, G. Vidal}
\affiliation{School of Mathematics and Physics, the University of Queensland, Brisbane 4072, Australia}
\date{\today}
 
\begin{abstract}
The ability of entanglement renormalization (ER) to generate a proper real-space renormalization group (RG) flow in extended quantum systems is analysed in the setting of harmonic lattice systems in $D=1$ and $D=2$ spatial dimensions. A conceptual overview of the steps involved in momentum-space RG is provided and contrasted against the equivalent steps in the real-space setting. The real-space RG flow, as generated by ER, is compared against the exact results from momentum-space RG, including an investigation of a critical fixed point and the effect of relevant and irrelevant perturbations. 
\end{abstract}

\pacs{05.30.-d, 02.70.-c, 03.67.Mn, 05.50.+q}

\maketitle 

\section{Introduction} \label{Sec:Intro}

The renormalization group (RG) is a set of tools and ideas used to investigate how the physics of an extended system changes with the scale of observation \cite{kadanov,wilson,fisher,cardy,shirkov,delamotte,shankar,dmrg,DMRGreview}. The RG plays a prominent role in the conceptual foundation of several areas of physics concerned with systems that consists of many interacting degrees of freedom, as is the case of quantum field theory, statistical mechanics and condensed matter theory \cite{kadanov,fisher,cardy,shirkov,delamotte,shankar}. In addition it also provides the basis for important numerical approaches to study such systems \cite{wilson,dmrg,DMRGreview,CORE1,CORE2}.

\begin{figure}[b]
  \begin{center}
    \includegraphics[width=8cm]{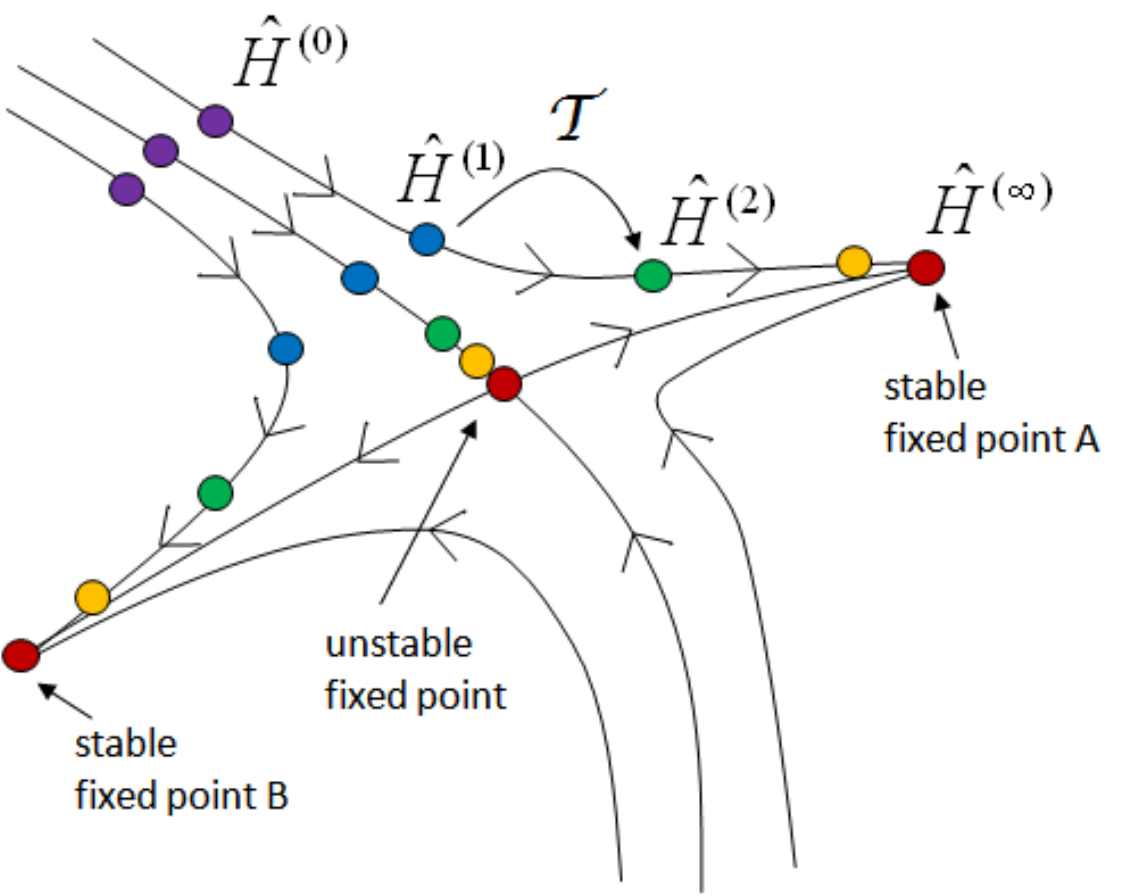}
  \caption{Given a description of an extended system in terms of a Hamiltonian $\hat H^{(0)}$ one may use successive RG transforms $\mathcal T$ to obtain a sequence of Hamiltonians $\left( \hat H^{(0)} , \hat H^{(1)} , \hat H^{(2)} , \ldots \right)$ each an effective theory describing the original system at successively lower energy scales (or longer distances). Characterising the \emph{fixed points} of the RG flow may help provide an understanding e.g. of the low-energy behavior of the system in the thermodynamic limit or of the stability of the system under various perturbations.}
  \label{fig:RGflow}
 \end{center}
\end{figure}

Given a microscopic description of an extended system in terms of its basic degrees of freedom and their interactions, RG methods aim to obtain an \emph{effective theory}, one that retains only some of these degrees of freedom but is nevertheless still able to reproduce its low energy (or long distance) physics. The effective theory is obtained through coarse-graining transformations that remove those degrees of freedom deemed to be \emph{frozen} at the observation scale of interest. For instance, given a Hamiltonian $\hat H^{(0)}$ for an extended system one may aim to use successive RG transforms to obtain a sequence of coarse-grained Hamiltonians $\left( \hat H^{(0)} , \hat H^{(1)} , \hat H^{(2)} , \ldots \right)$ each an effective theory describing the original system at successively lower energy scales, c.f. Fig. \ref{fig:RGflow}. Broadly speaking, RG techniques fall into two categories depending on how the coarse-graining is implemented, namely momentum-space RG \cite{delamotte} and real-space RG \cite{dmrg,DMRGreview,CORE1,CORE2}. 

Momentum-space RG is applied to theories that are expressed in Fourier space. It works by integrating out high-momentum modes of a field and it is often associated to perturbative approaches. Instead, real-space RG is applied directly to theories that are written in terms of local degrees of freedom, say spins in the case of a spin system defined on a lattice. It is not linked to perturbation theory and can in particular be applied to strongly interacting systems. As proposed by Kadanov \cite{kadanov}, the coarse-graining transformation is implemented by replacing a block of spins with a single effective spin, a procedure refined by Wilson \cite{wilson} and subsequently turned by White into the density matrix renormalization group (DMRG) algorithm \cite{dmrg, DMRGreview}, an impressively precise numerical tool to study one-dimensional systems.

A major difficulty of momentum-space RG comes precisely from the fact that it requires, as a starting point, a description of the system in Fourier space. Such description is not always available and might not be obtained easily. Consider for instance a system of interacting spins, as specified by some generic spin-spin interaction. There, obtaining a momentum-space representation might be as difficult as solving the whole theory. In this and many other cases, a RG approach must be performed in real space.

In spite of its indisputable success, the DMRG algorithm \cite{dmrg, DMRGreview} suffers from a shortcoming that has important implications. Because of the accumulation of short-ranged entanglement near the boundary of a spin block, the dimension of the Hilbert space used to effectively describe the block must grow with each iteration of the RG transformation. As a result, for instance, unstable fixed points of the RG flow (scale invariant critical systems) cannot be fixed points of the DMRG algorithm. Another, more practical consequence of this growth is that it limits the size of $1D$ critical systems that can be analyzed and, most importantly, it severely limits the success of DMRG computations in higher spatial dimensions. 

Entanglement renormalization (ER) is a real-space RG method recently proposed in order to overcome the above difficulties \cite{ER}. The main feature of ER is the use of \emph{disentanglers}. These are unitary transformations, locally applied near the boundary of a spin block, that remove short-ranged entanglement before the system is coarse-grained. As a result, the effective dimension of the Hilbert space for a spin block can be kept constant under successive RG transformations, so that the approach can be applied to arbitrarily large systems. The potential of ER ---as well as that of the related variational ansatz, the \emph{multi-scale entanglement renormalization ansatz} (MERA) \cite{MERA}--- to efficiently describe critical and non-critical ground states has been demonstrated for a number of spin and fermions models in one \cite{ER, unified, tMERA, algorithms,Transfer, CFT, CFT2, swingle} and two spatial dimensions \cite{evenbly,cincio,ER2D,ERKag,Ferm2,Ferm3,Ferm4,Ferm5}. Most significant are the recent works in which ER has been applied to study a geometrically frustrated spin model \cite{ERKag} and systems interacting fermions \cite{Ferm2, Ferm3, Ferm4,Ferm5}; these are problems that cannot be addressed by quantum Monte Carlo techniques due to the sign problem. In addition, it has been shown that the MERA offers a natural representation for systems with topological order \cite{QuantumDouble, StringNet}. Finally, several algorithms to compute the MERA have been put forward \cite{unified, tMERA, algorithms}.

In this work we explore the ability of entanglement renormalization to produce a sensible RG flow, one with the expected structure of fixed points and flow directions according to momentum-space RG, in $D=1,2$ dimensional harmonic lattice systems. Such systems are an ideal testing-ground for ER. On the one hand, they have well studied properties \cite{plenio, audenaert, skrovseth, cramer} and can be fully characterized in terms of correlation matrices, a fact that simplifies the analysis and conveniently reduces the computational complexity of ER calculations. On the other hand, an RG analysis of free-particle theories can be conducted simply and without approximations in momentum-space allowing for a comparison between the numerical results obtained using ER and the exact solution. The setting of harmonic lattices also allows for ER to be formulated in the language of bosonic modes, a formalism familiar to researchers in the areas of condensed matter physics and quantum field theory.

The results of Sect. \ref{Sec:Results} demonstrate that a real-space RG transform based upon entanglement renormalization is able to reproduce the exact results from momentum-space RG to a high accuracy in $D=1,2$ dimensional lattice systems, both for the critical and non-critical cases considered. Also demonstrated is the ability of the MERA to provide an efficient and accurate representation of the ground state of free boson systems, thus extending to the bosonic case the results of Ref. \cite{evenbly}. These results provide strong evidence that the ER approach, which can be implemented without making use of the special properties of free-particle systems, could be used to investigate low-energy properties of strongly interacting systems not tractable with momentum-space RG approaches \cite{unified, tMERA, algorithms}.

\begin{figure}[!tb]
  \begin{center}
    \includegraphics[width=8cm]{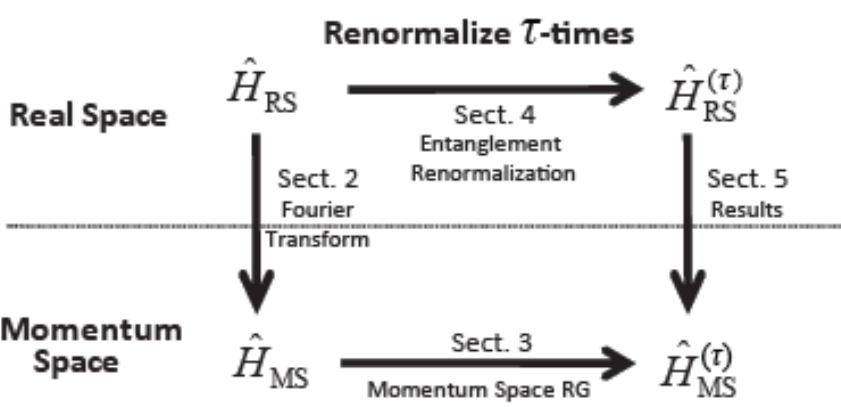}
  \caption{An outline of this paper. The harmonic lattice Hamiltonian $\hat H_{RS}$, defined in terms of interaction between local degrees of freedom, may be coarse-grained directly with a numeric implementation of a real-space RG transform, such as entanglement renormalization (ER), as outlined in Sect. \ref{Sec:RSRG}. Iterating the RG transform $\tau$ times we get the $\tau^\textrm{th}$ effective Hamiltonian $\hat H_{RS}^{(\tau)}$. An effective Hamiltonian may also be obtained by first transforming the Hamiltonian to a momentum-space representation $\hat H_\textrm{MS}$, via Fourier transform of the canonical coordinates, as described in Sect. \ref{Sec:Harm}. Momentum-space RG transforms may be applied (analytically) to $\hat H_\textrm{MS}$ as described Sect. \ref{Sec:MRG}. The dispersion relations from the real-space Hamiltonians $\hat H_{RS}^{(\tau)}$ are compared to those of the corresponding momentum-space Hamiltonians, $\hat H_\textrm{MS}^{(\tau)}$ in the results of Sect. \ref{Sec:Results}.}
  \label{ER-MSschematic}
 \end{center}
\end{figure}

The paper is organized in sections as follows (see also Fig. \ref{ER-MSschematic}). Sect. \ref{Sec:Harm} introduces the harmonic systems under consideration. In Sect. \ref{Sec:MRG} the process of renormalizing the system in momentum-space is explained, highlighting conceptual features of the RG. The critical system and examples of relevant and irrelevant perturbations are considered. Sect. \ref{Sec:RSRG} explains the details of the real-space RG implementation, both in terms of renormalizing the Hamiltonian and in terms of renormalizing the ground state directly. In Sect. \ref{Sec:Results} a comparison between results obtained from exact momentum-space RG and the results from numerical real-space RG is presented. 

\begin{figure}[tb]
  \begin{center}
    \includegraphics[width=8cm]{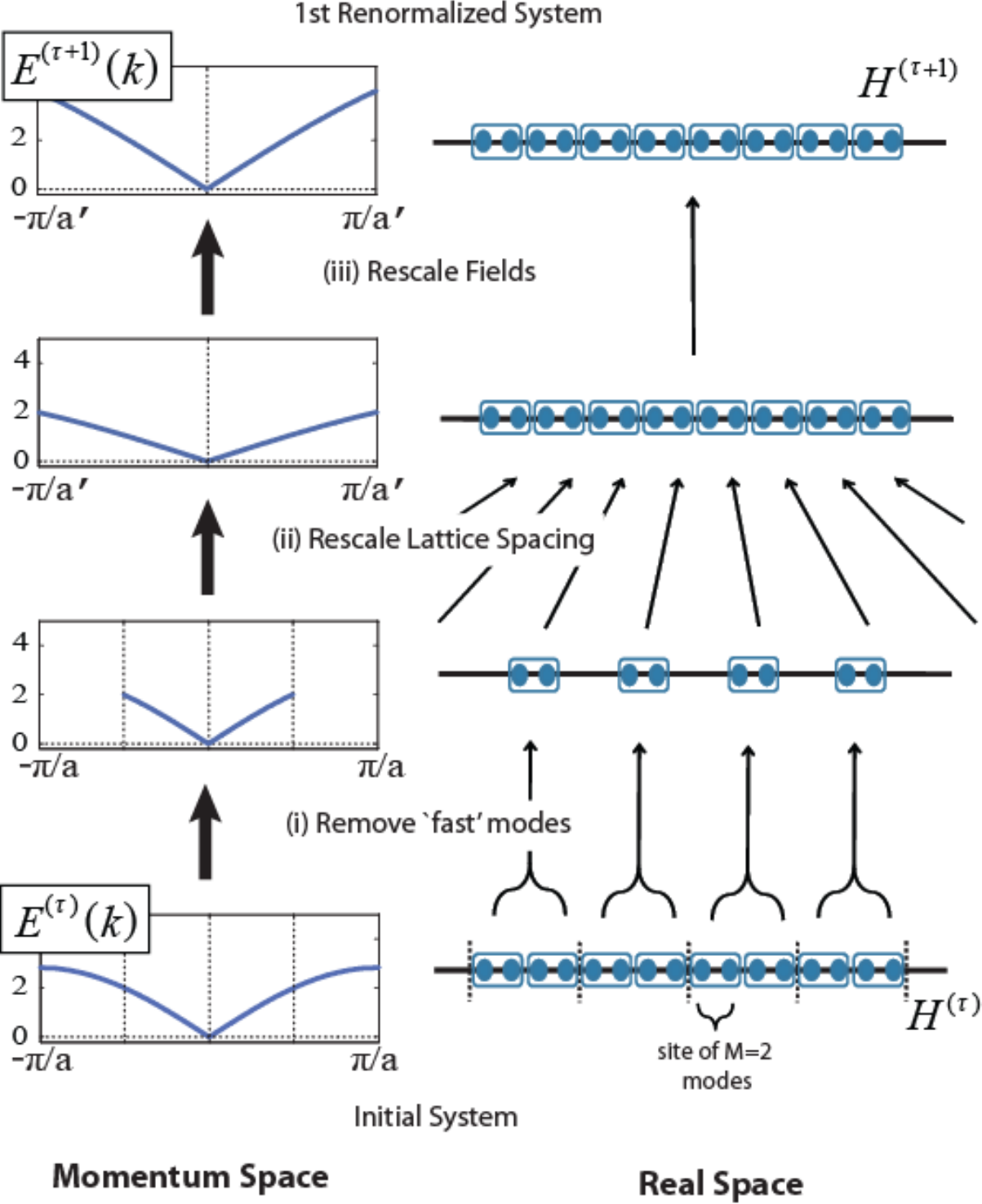}
  \caption{A comparison of an RG iteration for a $D=1$ dimensional system (left) in terms of a dispersion relation, $E(k)$, in momentum-space and (right) in terms of a lattice in real-space. The RG transformation maps the $\tau^\textrm{th}$ effective theory $\hat H^{(\tau)}$ into a new effective theory $\hat H^{(\tau+1)}$ while preserving the low-energy (or long distance) physics of the original theory. The three steps which comprise the iteration are described in detail in Sect. \ref{Sec:MRG} for momentum-space RG and Sect. \ref{Sec:HamRG} for the real-space RG. There are many possible ways the real-space coarse-graining step (i) can be implemented, Fig. \ref{ProjMera} describes two non-equivalent implementations.}
  \label{ERcompare}
 \end{center}
\end{figure}

\section{Coupled Harmonic Oscillators} \label{Sec:Harm}
In this work the low-energy subspaces of harmonic lattices in $D=1,2$ spatial dimensions are to be analysed using both real-space and momentum-space RG transformations. We begin with a brief introduction to harmonic lattices detailing equivalent representations in real-space and momentum-space coordinates, and the Fourier transform that shifts between such descriptions. For clarity the following derivations shall only be presented for the $1D$ system as the generalization to $2D$, or higher dimensional systems, is straight forward. The Hamiltonian for a chain of $N$ harmonic oscillators each with mass $m$, angular frequency $\omega$, and coupled with nearest neighbors via `spring constant' $K$, is written
\begin{align}
 \hat H &= c_0 \sum\limits_{r = 1}^N \left( {\frac{1}{{2m}}\hat p_r^2  + \frac{{m\omega ^2 }}{2}\hat q_r^2  + K\left( {\hat q_{r + 1}  - \hat q_r } \right)^2 }\right)   \nonumber\\ 
  &= \sum\limits_{r = 1}^N \left( {\hat p_r^2  + m^2 \omega ^2 \hat q_r^2  + 2\tilde K\left( {\hat q_{r + 1}  - \hat q_r } \right)^2 }\right) \label{s1e1}   
\end{align}
where in the second line we have chosen $c_0=2m$ and defined $\tilde K=mK$ for convenience. Note that periodic boundary conditions are assumed. The operators $\hat p_i$ and $\hat q_i$ are the usual canonical coordinates with commutation $\left[ {\hat p_k ,\hat q_l } \right] = i\hbar\delta_{kl}$. In our present considerations it is convenient to focus on the critical (massless) Hamiltonian
\begin{equation}
\hat H_0  = \sum\limits_{r = 1}^N \left( {\hat p_r^2  + 2\tilde K\left( {\hat q_{r + 1}  - \hat q_r } \right)^2 } \right), \label{s1e2}
\end{equation}
though the non-zero mass case will later be reintroduced in Sect. \ref{Sec:Rel}. As a preliminary to the momentum-space RG, the Hamiltonian $\hat H_0$ shall be recast into momentum-space variables via Fourier transform of the canonical coordinates. The Fourier-space coordinates $\check p$ and $\check q$ are defined
\begin{align}
 \check p_\kappa   = \frac{1}{{\sqrt N }}\sum\limits_{r = 1}^N {\hat p_r e^{ - 2\pi ir\kappa /N} }  \nonumber\\ 
 \check q_\kappa   = \frac{1}{{\sqrt N }}\sum\limits_{r = 1}^N {\hat q_r e^{ - 2\pi ir\kappa /N} } .  \label{s1e2b}
\end{align}
Substitution of the Fourier-space coordinates brings the Hamiltonian of Eq. \ref{s1e2} into diagonal form, here a set of $N$ uncoupled oscillators
\begin{equation}
\hat H_0  = \sum\limits_{\kappa  =  - (N - 1)/2}^{(N - 1)/2} \left( {\check p_\kappa ^2  + 8\tilde K\sin ^2 \left( \frac{\pi \kappa}{ N} \right)\check q_\kappa ^2 }\right) .\label{s1e3} 
\end{equation}
Defining $k=2\pi\kappa/(aN)$, with constant `$a$' representative of the lattice spacing, the thermodynamic limit ($N\rightarrow \infty$) can be taken
\begin{equation}
\hat H_0  = \int\limits_{k =  - \pi/a }^{\pi/a}  \left( {\check p\left( k \right)^2  + 8\tilde K\sin ^2 \left( \frac{ka}{2} \right)\check q\left( k \right)^2 }\right)dk . \label{s1e4}
\end{equation}
The theory has a natural momentum cut-off $\Lambda=\pm \pi/a$ originating from the finite lattice spacing; taking the lattice spacing `$a$' to zero recovers the continuum limit with corresponds to the field theory of a real, massless scalar field. Two equivalent representations of the harmonic chain have been obtained; that of Eq. \ref{s1e2} written in terms of spatial modes (amenable to numeric, real-space RG) and that of Eq. \ref{s1e4} written in terms of momentum modes (amenable to analytic, momentum-space RG). The \emph{dispersion relation} $E_0(k)$ of the system, which describes the energy of momentum mode $k$ and is known from the solution to a single oscillator, is given by
\begin{equation}
E_0 \left( k \right) = 2\sqrt 2 \tilde K\left| {\sin \left( {ka/2} \right)} \right|. \label{s1e5}
\end{equation}
The RG transformations of the Hamiltonian shall be chosen such that the resulting \emph{effective theory} preserves the low-energy structure of the original theory, or equivalently, preserves the small $k$ part of the dispersion relation. As the low-energy part of the dispersion is gapless and linear we would expect that, under the RG flow, the renormalized Hamiltonians would tend to a fixed point that has a linear, gapless dispersion.

\section{Momentum-Space RG} \label{Sec:MRG}
In this section the harmonic system of Eq. \ref{s1e4}, which has been recast in a Fourier basis, is analysed using momentum-space RG. Although the RG analysis of interacting systems is more complicated that the free particle analysis undertaken here, and often involves perturbation theory, the basic procedure is the same. It is useful to consider the RG transformation as occuring in three steps. (i) Firstly the momentum cut-off is reduced, $\Lambda\mapsto \Lambda'=\Lambda/2$, and modes greater than the cut-off are integrated out of the theory. As there is no interaction between the momentum modes of Eq. \ref{s1e4}, this step is presently very simple; the cut-off is reduced $\Lambda\mapsto\Lambda'=\Lambda/2$ whilst leaving the form of the Hamiltonian for modes with momentum $k<\Lambda'$ unchanged
\begin{equation}
\hat H'_0  = \int\limits_{k =  - \pi /2a}^{\pi /2a} \left( {\check p\left( k \right)^2  + 8\tilde K\sin ^2 \left( \frac{ka}{2} \right)\check q\left( k \right)^2 }\right) dk. \label{s2e1}
\end{equation}
(ii) Next the length associated to the system is changed, this can be implemented as a scaling of the lattice spacing\footnote{The approach of rescaling the lattice spacing is common in the context of condensed matter problems; equivalently we could have rescaled the momentum of the theory, $k\mapsto k'=2k$, as is the approach most often used for the RG in a quantum field theory setting.}, $a\mapsto a'=2a$, which gives
\begin{equation}
\hat H''_0  = \int\limits_{k =  - \pi/a' }^{\pi/a'} \left( {\check p\left( {k} \right)^2  + 8\tilde K\sin ^2 \left( \frac{ka'}{4} \right)\check q\left( {k} \right)^2 }\right) dk. \label{s2e2}
\end{equation}
A change has been made to the observation scale of the system in terms of \emph{length}. Next a change in the observation scale is made in terms of \emph{energy}. Indeed, in the final step (iii) the fields are rescaled
\begin{align}
 \check p\left( {k} \right) \mapsto \check p'\left( {k} \right) &= \frac{1}{\sqrt{2 }}\check p\left( {k} \right) \nonumber\\ 
 \check q\left( {k} \right) \mapsto \check q'\left( {k} \right) &= \frac{1}{\sqrt{2 }}\check q\left( {k} \right) \label{s2e3},
\end{align}
so that the new field operators have a modified commutation relation $[p'(k),q'(k)]=i\hbar/2$, in accordance with the desired change of energy scale. In principle the RG transformation is complete, however in this instance a further transform is required to recast the critical Hamiltonian into a manifestly invariant form. The field operators are rescaled once more 
\begin{align}
 \hat p' \mapsto \hat p'' &= \sqrt{2} \hat p' \nonumber\\ 
 \hat q' \mapsto \hat q'' &= \frac{1}{{\sqrt{2}}}\hat q'.  \label{s2e4}
\end{align}
In contrast to the previous transform of Eq. \ref{s2e3}, this transform is \emph{commutation preserving} hence does not affect the physics of the system; namely the dispersion relation remains unchanged. Implementing the third step, together with the auxiliary transform of Eq. \ref{s2e4}, the first renormalized Hamiltonian $\hat H_0^{(1)}$ is given
\begin{equation}
\hat H_0^{(1)}  = \int\limits_{k =  - \pi/a' }^{\pi/a'}  \left( {\check p''\left( {k} \right)^2  + 32\tilde K\sin ^2 \left( \frac{ka'}{4} \right)\check q''\left( {k} \right)^2 }\right) dk. \label{s2e5}
\end{equation}
The RG transformation is summarized: (i) the degrees of freedom that are not relevant to the low energy physics are removed, followed by a changes of observation scale in terms of (ii) \emph{length} ($a\mapsto a'=2a$) and (iii) \emph{energy} ($\hbar\mapsto \hbar'=\hbar/2$). Starting from a theory with a natural length scale, the lattice spacing $a$, the RG transform is thus used to derive an effective theory with the new length scale $a'$. The scale factors chosen in steps (ii) and (iii) may depend on the implementation of the RG as well as the problem to which it is being applied. Iterating the RG transform (dropping the `primes' from notation), the $\tau^\textrm{th}$ renormalized Hamiltonian is 
\begin{equation}
 \hat H_0^{(\tau)}  = \int\limits_{k =  - \pi/a }^{\pi/a}  \left( {\check p\left( k \right)^2  + 2^{2\tau + 3} \tilde K\sin ^2 \left( \frac{ka}{2^{\tau + 1} } \right)\check q\left( k \right)^2 }\right) dk,  \label{s2e6a}
\end{equation}
with corresponding dispersion relation
\begin{equation}
E_0^{(\tau)} \left( k \right) = 2^{\tau + \frac{3}{2}} \sqrt{\tilde K}\left| {\sin \left( {ka/2^{\tau + 1} } \right)} \right| .\label{s2e6b}
\end{equation}
In the limit of infinitely many transforms, $\tau\rightarrow\infty$, we get the Hamiltonian $H_0^{(\infty)}$ at the fixed point of the RG flow 
\begin{equation}
H_0^{(\infty)} = \int\limits_{k =  - \pi/a }^{\pi/a}  \left( \check p\left( k \right)^2  + \tilde K \left( k^2 a^2 \right)\check q\left( k \right)^2 \right) dk \label{s2e6}. 
\end{equation}
The fixed point Hamiltonian has a purely linear, gapless dispersion
\begin{equation}
  E_0^{(\infty)}\left( k \right)= a\sqrt {2 \tilde K} \left| k \right|  .\label{s2e7} 
\end{equation}
as anticipated.

\subsection{Relevant perturbation} \label{Sec:Rel}
In the previous section the massless harmonic lattice of Eq. \ref{s1e4}, a critical system, was shown to be a (non-trivial) fixed point of the RG flow. We now consider the stability of this Hamiltonian under the addition of perturbations. Perturbations to the critical theory can be classified as being relevant or irrelevant depending on whether the deviations from the fixed point, induced by the perturbations, grow or diminish under the RG flow. In order to study relevent perturbations a mass term $\hat H_{{\rm{rel}}}$ is reintroduced to the critical system of Eq. \ref{s1e4}; as shall be shown shortly this term \emph{grows} under the RG flow. Hence, it significantly modifies the low-energy physics from that of the unperturbed system. The mass term is diagonal in both real-space and momentum-space representations
\begin{equation}
 \hat H_{{\rm{rel}}} \equiv \sum\limits_{r} {\hat q_r^2 } = \int\limits_{k =  - \pi/a }^{\pi/a}  { \check q\left( k \right)^2 dk}. \label{s3e1}  
\end{equation}
The perturbed Hamiltonian $\hat H_m = \hat H_0 + m^2 \hat H_\textrm{rel}$ is equal to the original Hamiltonian of Eq. \ref{s1e1} describing coupled harmonic oscillators of mass $m$. Since this perturbation term does not reorder mode energies ($E(k)$ is still an increasing function of $|k|$) the same RG transformations may be performed on $\hat H_\textrm{rel}$ as on the unperturbed system $\hat H_0$. This analysis shows that the $\tau^\textrm{th}$ renormalized perturbation term $\hat H_\textrm{rel}^{(\tau)}$ grows exponentially with the RG iteration $\tau$
\begin{equation}
\hat H_\textrm{rel}^{(\tau)}  = 2^{2\tau} \hat H_\textrm{rel},\label{s3e2} 
\end{equation}
Thus even the addition of even a small mass $m$ to the critical system leads to a large difference between the perturbed $\hat H_m$ and original $\hat H_0$ Hamiltonians after only a few RG iterations; that is to say, the perturbed system tends to a different fixed point of the RG flow. This is further evidenced by consideration of the dispersion relation $E_m^{(\tau)}$ of the perturbed system $\hat H_m$
\begin{align}
 E_m^{(\tau)} \left( k \right) &= 2^\tau \sqrt {m^2  + 8\tilde K\sin ^2 \left( {ka/2^{\tau + 1} } \right)} \nonumber \\ 
  &= 2^\tau m + \frac{{a^2\tilde K}}{{2^\tau m}}k^2 + O\left( {\frac{a^2 \tilde K}{{m^3  2^{3\tau -3} }}} \right).\label{s3e3}
\end{align}
In contrast to the linear, gapless dispersion $E_0^{(\tau)}$ of the massless theory described by Eq. \ref{s2e7}, we now see a quadratic dependence of the energy on the momentum $k$, together with an energy gap that grows \emph{exponentially} with the number $\tau$ of RG iterations.

\subsection{Irrelevant perturbation} \label{Sec:Irrel}
Perturbations that become smaller along the RG flow are termed \emph{irrelevant perturbations}, as these do not affect the asymptotic, low-energy behavior of the system. In this section we construct and analyze an example of such a perturbation. The irrelevant perturbation $\hat H_{{\rm{irrel}}}$ is constructed from neighboring and next-to-nearest neighboring quadratic couplings
\begin{equation}
 \hat H_{{\rm{irrel}}}  \equiv \sum\limits_{r = 1}^N \left( { - \hat q_r^2  + \left( {\hat q_{r + 1}  - \hat q_r } \right)^2  + \frac{1}{4}} \left( {\hat q_{r + 2}  + \hat q_r } \right)^2\right). \label{s4e1}
\end{equation}
In the Fourier basis $\hat H_{{\rm{irrel}}}$ has representation
\begin{equation}
\hat H_{{\rm{irrel}}} = 4 \int\limits_{k =  - \pi/a }^{\pi/a} {\sin ^4}\left( \frac{ ka}{2}\right) \check q\left( k \right)^2 dk. \label{s4e1b}
\end{equation} 
We consider the perturbed system $\hat H_\alpha= \hat H_0 +\alpha \hat H_\textrm{irrel}$ with the perturbation strength chosen $\alpha>0$. The $\tau^\textrm{th}$ renormalized perturbation $\hat H_{{\rm{irrel}}}^{(\tau)}$ may be obtained through the same sequence of RG transforms as applied to the unperturbed system
\begin{align}
\hat H_\textrm{irrel}^{(\tau)}  &= \int\limits_{k =  - \pi/a }^{\pi/a}  {2^{2\tau + 2} \sin ^4 \left( \frac{ka}{2^{\tau + 1}} \right)\hat q\left( k \right)^2 dk}  \nonumber\\ 
  &= 2^{ - 2\tau - 2} \int\limits_{k =  - \pi/a }^{\pi/a}  {\left(a^4 k^4+ O\left(\frac{a^6 k^6}{2^{2\tau + 2}}\right) \right) \hat q\left( k \right)^2 dk},  \label{s4e2} 
\end{align}
The perturbation $\hat H_\textrm{irrel}$ is thus exponentially suppressed under the RG flow. Equivalently, the addition of this term to the critical system has an effect on the low-energy physics that diminishes with each RG iteration, as is seen directly from the dispersion $E_\alpha^{(\tau)}$ of the perturbed system $\hat H_\alpha$
\begin{align}
 E_{\alpha}^{(\tau)} (k) &= 2^{\tau + 1} \left| {\sin \left( \frac{ka}{2^{\tau + 1} } \right)} \right|\sqrt {2\tilde K + \alpha \sin ^2 \left( \frac{ka}{2^{\tau + 1} } \right)} \nonumber\\ 
  &= a\sqrt {2\tilde K} \left| k \right| + O\left( {2^{ - 2\tau} } \right),  \label{s4e3}
\end{align}
which converges to the same linear, gapless fixed point as did the unperturbed system of Eq. \ref{s2e7}. 

\begin{figure}[!tb]
  \begin{center}
    \includegraphics[width=8cm]{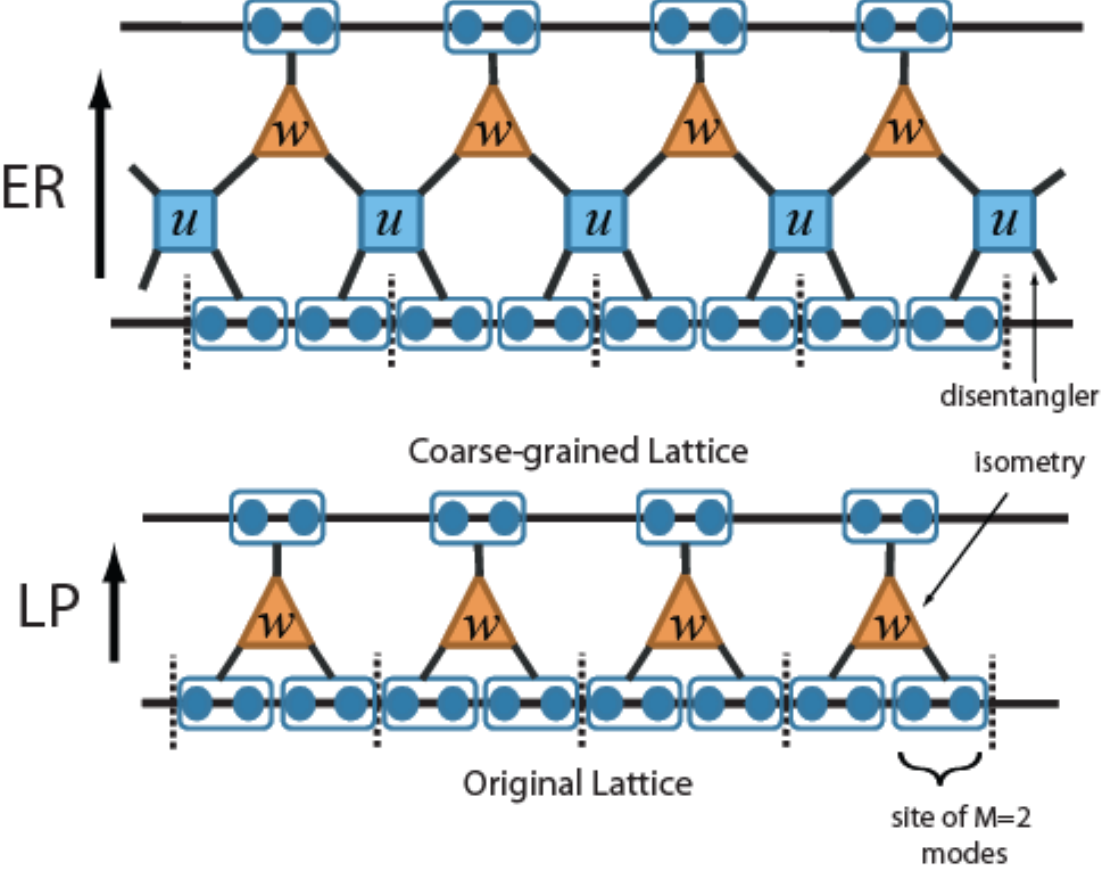}
  \caption{The coarse-graining step of the real-space RG, in which a block of two sites (each composed of $M$ bosonic modes) of the original lattice $\mathcal L$ is mapped to a single site of the coarse-grained lattice $\mathcal L '$, can be accomplished in different ways. (Bottom) The simplest method involves applying a series of local projections, realized by isometric tensors $w$, whose function is to select and retain relevant block degrees of freedom. This shall subsequently be referred to as the local projection (LP) method. (Top) Entanglement Renormalization (ER) differs from the LP coarse-graining by the inclusion unitary disentanglers $u$, enacted across block boundaries, before the truncation with isometries.}
  \label{ProjMera}
 \end{center}
\end{figure}

\section{Real-Space RG} \label{Sec:RSRG}
In this section an implementation of real-space RG based upon coarse-graining transformations of the lattice is described. Following the seminal works of Migdal, Kadanoff, and Wilson in real-space RG \cite{kadanov, wilson}, the coarse-graining transformation maps a \emph{block} of sites from the original lattice $\mathcal L$ into a single site of a coarser lattice $\mathcal{L}'$. Let us consider a $1D$ lattice $\mathcal{L}$ of $N$ sites, each site described by a vector space $\mathbb V$. We divide $\mathcal L$ into blocks of two sites and, following Wilson, implement a coarse-graining transformation by means of an isometry $w$
\begin{equation}
w: \mathbb V' \mapsto \mathbb V^{\otimes 2},\ \ w^{\dag} w =I_\mathbb{V'} \label{s5e1}
\end{equation}
where $\mathbb V^{\otimes 2}$ is the vector space of two sites, $\mathbb V'$ is the vector space of a site in the coarser lattice $\mathcal L'$ of $N'=N/2$ sites and $I_\mathbb{V'}$ is the identity in $\mathbb V'$. This coarse-graining transformation, which we shall refer to as a \emph{local projection} (LP) transformation, is depicted graphically in Fig. \ref{ProjMera}. From an initial Hamiltonian $\hat H$ defined on lattice $\mathcal L$ we can obtain an effective Hamiltonian $\hat H'$ on lattice $\mathcal L'$ via the transformation
\begin{equation}
\hat H'=W^\dag \hat H W,\ \  W = w^{\otimes N/2}. \label{s5e2}
\end{equation}
The LP transformation has the property of preserving locality of operators; for instance if the original Hamiltonian was a sum of two-body interactions, $\hat H = \sum\nolimits_{i = 1}^N {h_{i,i + 1} }$, then the effective Hamiltonian would remain a sum of two-body interactions, $\hat H' = \sum\nolimits_{i = 1}^{N'} {h'_{i,i + 1} }$. An alternative coarse-graining transformation, known as entanglement renormalization (ER), follows as a modification of the LP scheme. As with the LP scheme, we map two sites of $\mathcal L$ into a single effective site of $\mathcal L'$ via an isometry $w$, however in ER one first enacts unitary \emph{disentanglers} $u$
\begin{equation}
u: \mathbb V^{\otimes 2} \mapsto \mathbb V^{\otimes 2},\ \  u^{\dag} u = u u^\dag =I_{\mathbb{V}^{\otimes 2}} \label{s5e3}
\end{equation}
across the boundaries of adjacent blocks, as show Fig. \ref{ProjMera}. Thus the effective Hamiltonian $\hat H'$, as given by a transformation with entanglement renormalization, is  
\begin{equation}
\hat H'=\left( W^\dag U^\dag \right) \hat H \left( U W \right) ,\ \  U = u^{\otimes N/2}. \label{s5e4}
\end{equation}
Inclusion of the disentanglers in the RG scheme, although having the unfortunate effect of increasing the computational cost of numeric implementation, has profound implications regarding the ability of the method to produce a sensible real-space RG flow, as shall be demonstrated in the results of Sect. \ref{Sec:Results}. The action of the disentanglers spreads local operators; if the original Hamiltonian was a sum of two-body interactions, $\hat H = \sum\nolimits_{i = 1}^N {h_{i,i + 1} }$, then the effective Hamiltonian would consist of three-body interactions, $\hat H' = \sum\nolimits_{i = 1}^{N'} {h'_{i,i + 1,i+2} }$. However, under further transformations the interactions remain fixed at three body interactions, a property of the ER transformation referred to as a bounded causal width \cite{MERA}. A more comprehensive introduction to real-space RG, including a discussion on the role of disentanglers and a description of ER schemes for $2D$ lattice systems can be found in Refs. \cite{ER, MERA, algorithms}. 

Starting from the original Hamiltonian $( \mathcal{L}^{(0)} ,\hat H^{(0)} ) \equiv ( \mathcal{L} ,\hat H )$ one could, by iteration of the RG map, generate a sequence of effective Hamiltonians defined on successively coarser lattices  
\begin{equation}
\left( \mathcal{L}^{(0)}, \hat H^{(0)} \right) \stackrel{\mathcal{T}_0}{\rightarrow} \left( \mathcal{L}^{(1)}, \hat H^{(1)} \right) \stackrel{\mathcal{T}_1}{\rightarrow} \left( \mathcal{L}^{(2)}, \hat H^{(2)} \right) \stackrel{\mathcal{T}_2}{\rightarrow} \ldots \label{s5e5},
\end{equation}
with the transformation $\mathcal T$ representing either a LP or ER transformation. The LP transformation $\mathcal T_\textrm{LP}^{(\tau)}$ is characterised by the corresponding isometry $w^{(\tau)}$, while the ER transformation $\mathcal T_\textrm{ER}^{(\tau)}$ is characterised jointly by an isometry and a disentangler, $(w^{(\tau)},u^{(\tau)})$. Equation \ref{s5e5} can be used to directly investigate how $\hat H$ changes under scale transformations allowing e.g. one to characterise the stability of $\hat H$ under perturbations or investigate properties of the system in the thermodynamic limit directly. 

In order to make meaningful comparisons between effective Hamiltonians defined at different length scales, say between $\hat H^{(\tau)}$ and $\hat H^{(\tau+1)}$ defined on lattices $\mathcal L^{(\tau)}$ and $\mathcal L^{(\tau+1)}$ respectively, it is required that the dimension of the Hilbert space of a site in $\mathcal L^{(\tau+1)}$ remains the same as that of $\mathcal L^{(\tau)}$, or equivalently that $\mathbb V' = \mathbb V$ in Eq. \ref{s5e1}. This ensures that, for instance, two-body operators defined on $\mathcal L^{(\tau)}$ and $\mathcal L^{(\tau+1)}$ exist in the same parameter space, allowing a direct comparison between the coefficients that describe the operators. Keeping the dimension of effective lattice sites in lattice $\mathcal L^{(\tau)}$ constant between RG transforms is also desirable for computational purposes. Indeed, if the dimension of effective sites were to grow with each RG iteration then the computational cost of implementing the transformations would also grow, limiting the number of transforms which may be performed. An investigation of whether the LP and ER transformations can accurately coarse-grain Hamiltonians over repeated RG iterations while keeping a fixed local dimension is a focus of this work.

Note that, in general, the disentanglers and isometries $(u,w)$ that best preserve the low-energy space of a Hamiltonian $\hat H^{(0)}$ under coarse-graining will depend on the specific Hamiltonian $\hat H^{(0)}$ itself; in practice optimisation techniques \cite{unified, tMERA, algorithms} are required to compute the tensors $(u,w)$. There are however some systems where analytic expressions for these tensors may be obtained \cite{QuantumDouble, StringNet}.

\subsection{RG of the Harmonic Lattice} \label{Sec:HamRG}
We now turn our attention to the analysis of the harmonic lattice system with ER\footnote{The LP coarse-graining may be viewed as a simplification of ER in which the disentanglers $u$ are set to identity transforms. Therefore only implementation of ER need be explicitly described.}. Following the ideas discussed in the previous section, several possible algorithms \cite{unified, tMERA, algorithms} could be applied to study the low-energy subspace of the harmonics systems directly. Application of these algorithms requires only that the Hamiltonian in question is composed of a sum of local interaction terms, as is the case with the systems we consider. However, as this study is focused on Hamiltonians containing only \emph{quadratic couplings}, this property can be exploited in order to significantly reduce the cost of the numerical RG as well as simplify the analysis of the results. The Hamiltonian of Eq. \ref{s1e1}, describing a $1D$ harmonic chain of $N$ modes, can be written in a concise quadratic form
\begin{equation}
 \hat H = \sum\limits_{i,j = 1}^{2N} { R_i^T \mathcal{H}_{ij}  R_j }\label{s6e1}   
\end{equation}
by defining a quadrature vector $\vec R \equiv \left( {\vec p,\vec q} \right)$ with
\begin{equation}
\vec p \equiv \left( {\begin{array}{*{20}c}
   {\hat p_1 }  \\
   {\hat p_2 }  \\
    \vdots   \\
   {\hat p_N }  \\
\end{array}} \right),\vec q \equiv \left( {\begin{array}{*{20}c}
   {\hat q_1 }  \\
   {\hat q_2 }  \\
    \vdots   \\
   {\hat q_N }  \\
\end{array}} \right)\label{s6e1b}
\end{equation}
where $\mathcal H$, henceforth referred to as the Hamiltonian matrix, is a $2N\times 2N$ Hermitian matrix. The coarse-graining transformations shall be chosen such that the effective Hamiltonians also only contain quadratic couplings, described by some new Hamiltonian matrix $\mathcal H'$. Thus the RG analysis may be performed in the space of Hamiltonian matrices $\mathcal H$, as opposed to the (much larger) space of full-fledged Hamiltonians $\hat H$ and a more efficient realization of ER is possible. Retaining the quadratic form of the Hamiltonian entails limiting the disentanglers $u$ and isometries $w$ which comprise the ER map to \emph{cannonical} transformations, namely transformations that preserve commutation relations. Consider a transformation of the Hamiltonian matrix by a $2N \times 2N$ matrix $S$
\begin{equation}
\mathcal{H} \mapsto \mathcal{H}' = S^T \mathcal{H}S.\label{s6e2}
\end{equation}
In order for the transformation to be commutation preserving it is required that the transform $S$ be a symplectic matrix, $S\in \textrm{Sp}(2N,\mathbb{R})$. A symplectic transform can be characterized as leaving the symplectic matrix $\Sigma$ invariant under conjugation, $S^T \Sigma S = \Sigma$. Given our convention of grouping the quadrature vectors in Eq. \ref{s6e1b} the symplectic matrix takes the form 
\begin{equation}
\Sigma  \equiv \left( {\begin{array}{*{20}c}
   0 & {{I}_N }  \\
   {{I}_N } & 0  \\
\end{array}} \right)\label{s6e2b}
\end{equation}
with $I_N$ as the $N\times N$ identity. Additionally, the Hamiltonians under consideration take even simpler form than Eq. \ref{s6e1}; as there is no coupling between $\hat p$ and $\hat q$ quadrature degrees of freedom in the harmonic chain, the Hamiltonian may be expressed as
\begin{equation}
\hat H = \vec p^T \vec p + \vec q^T \mathcal{H}_q \vec q. \label{s6e3} 
\end{equation}
It is thus convenient to restrict symplectic transforms $S$ to those which preserve the form of Eq. \ref{s6e3}; we only consider transforms of the type $S =V  \oplus V$, with $V$ a \emph{special orthogonal} transformation, $V\in \textrm{SO}(N)$. It can be easily checked that $V  \oplus V$ is a symplectic transform. In fact, this is an element of the maximal compact subgroup of Sp$(2N,\mathbb R)$. The $\hat p$-quadrature part of the Hamiltonian in Eq. \ref{s6e3} remains trivial under these transformations, allowing us to focus on the $\hat q$-quadrature part $\mathcal H_q$ of the Hamiltonian matrix, which transforms as
\begin{equation}
{\mathcal{H}_q}' = V^T \mathcal{H}_q V.  \label{s6e4}
\end{equation}
Let us group a number $M$ of contiguous bosonic modes of the $1D$ harmonic chain together; each group of $M$ modes shall henceforth be referred to as a \emph{site} of the original lattice $\mathcal{L}$. The disentanglers $u$, which act on two sites (that is, on $2M$ modes), are chosen as special orthogonal transforms $u\in \textrm{SO}(2M)$. Isometries $w$ are realized as a composition of a special orthogonal transform followed by a projection, $w = w_0 w_\textrm{proj.}$, with
\begin{equation}
w_0 \in \textrm{SO}(2M),\ \  w_\textrm{proj.} = (0_M  \oplus I_M).  \label{s6e5}
\end{equation}
The transformation of the entire lattice is achieved by first constructing the direct sum of the local operators
\begin{equation}
W =  {\mathop  \bigoplus \limits_{i = 1}^{N/2M} w},\ \   U =  {\mathop  \bigoplus \limits_{i = 1}^{N/2M} u}.  \label{s6e6}
\end{equation}
Given the disentangler and isometry $(u,w)$, the Hamiltonian $\mathcal H_q$ is coarse-grained into a new Hamiltonian $\mathcal H_q '$ defined as
\begin{equation}
{{\mathcal{H}}_q} '= W^\textrm{T} U^\textrm{T} \left( \mathcal{H}_q \right) U W.\label{s6e7}
\end{equation}
By the definition of the isometry in Eq. \ref{s6e5} it is ensured that if the initial lattice $\mathcal L$ has $M$ bosonic modes per lattice sites then the coarser lattices $\mathcal L ^{(\tau)}$ also have $M$ modes per lattice site. As discussed earlier in Sect. \ref{Sec:RSRG}, keeping the number of degrees of freedom per lattice site constant between RG maps is necessary to allow meaningful comparison of operators at different length scales. Note that the number of modes per site $M$ plays the role of a refinement parameter; choice of a larger $M$ retains more parameters in the description of the effective theory, yielding more accurate results, at the cost of greater computational expense. The simplest choice of a one-to-one correspondence between bosonic modes and lattice sites, i.e. setting $M=1$, does not give sufficiently accurate numerics, hence the need for grouping $M>1$ modes into each lattice site. 

\begin{figure}[!tb]
  \begin{center}
    \includegraphics[width=8cm]{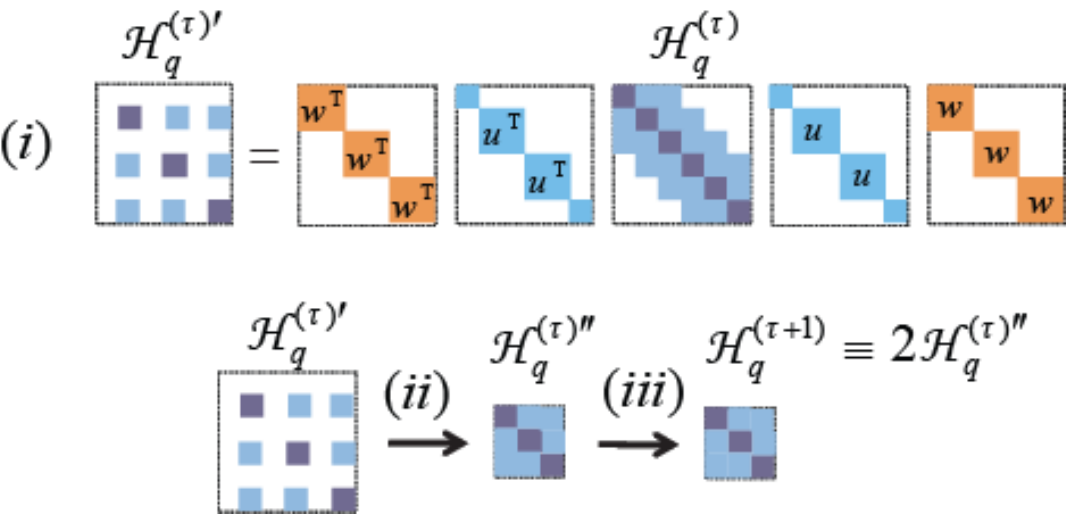}
  \caption{An iteration of entanglement renormalization, broken into three steps, is depicted in terms of the direct sum structure of the Hamiltonian matrix $\mathcal H_q$. Dark shaded squares in $\mathcal H_q$ represent couplings within the site of $M$ modes, light shaded squares are the couplings between sites (at most next-to-nearest neighbor). Step(i), removing `fast' modes is realized by transforming $\mathcal H_q$ by conjugation with the disentanglers $u$ and isometries $w$. Step(ii), rescaling the momentum, is achieved by removing the zero rows/columns from $\mathcal H_q'$. Step(iii), rescaling the fields, is achieved by directly scaling $\mathcal H_q''$ by a factor of 2. These three steps combined take the $\tau^\textrm{th}$ renormalized Hamiltonian matrix $\mathcal H_q^{(\tau)}$ to the $(\tau+1)^\textrm{th}$ renormalized Hamiltonian matrix $\mathcal H_q^{(\tau+1)}$. }
  \label{RScovariance}
 \end{center}
\end{figure}

It is only for a proper choice of disentangler and isometry $(u,w)$ that $\mathcal H_q '$ of Eq. \ref{s6e7} retains the low-energy subspace of the original $\mathcal H_q$. This proper choice is found by optimisation over all possible $(u,w)$. It is desired that the RG transform be optimized to project onto the \emph{minimum} energy subspace of the original Hamiltonian; a matrix equation which describes the minimization can be written
\begin{equation}
\mathop {\min }\limits_{u,w} \left( \textrm{tr} \{ {H_q '} \} \right),  \label{s6e8}
\end{equation}
with the effective Hamiltonian $\mathcal H_q '$ as Eq. \ref{s6e7}. The matrix $\mathcal H_q '$ describes a Hamiltonian that is translation invariant between blocks of $2M$ modes, hence the trace of $\mathcal H_q '$ (which describes the entire system) may be minimised by minimising the trace of an individual block of $\mathcal H_q '$. An optimisation method, based upon alternating updates for isometries $w$ and disentanglers $u$, can be used to find suitable $(u,w)$ that minimise Eq. \ref{s6e8} and best preserve the low-energy space of the original Hamiltonian $\mathcal H_q$. The optimisation method to be used here is similar to the general algorithm described in Sect. IV of Ref. \cite{algorithms}.

Assuming that suitable $(u,w)$ have been obtained, the full real-space RG transformation of the Hamiltonian may be achieved in three steps, as illustrated Fig. \ref{RScovariance}, analogous to the three steps of momentum-space procedure described Sect. \ref{Sec:MRG}. Firstly, (i) the Hamiltonian matrix $\mathcal H_q$ is transformed with direct sums of the disentanglers and isometries as Eq. \ref{s6e7}. Recall from Eq. \ref{s6e5} that an isometry $w$, which acts upon a block of $2M$ modes, consists of a special orthogonal transform followed by a projection, $w = w_0 w_\textrm{proj}$. The projection has form $w_\textrm{proj}=(0_M\oplus I_M)$ with the trivial (zero) part $0_M$ describing the $M$ modes to be removed from the system and the identity part $I_M$ describing the $M$ modes retained in the effective description. In the next step, (ii) the zero rows/columns, those which were acted upon in the previous step by $0_M$, are removed from Hamiltonian matrix to form a new matrix ${\mathcal{H}_q^{(0)}} ''$. (iii) The final step of rescaling the fields is realized by scaling the Hamiltonian matrix by a factor, the same rescaling as Eqs. \ref{s2e3} and \ref{s2e4} for momentum-space RG is realized by defining $\mathcal{H}_q^{(1)}\equiv 2 {\mathcal{H}_q^{(0)}} ''$, with ${\mathcal{H}_q^{(1)}}$ as the first renormalized Hamiltonian. Iterating this procedure $\tau$ times gives the $\tau^\textrm{th}$ renormalized Hamiltonian ${\mathcal{H}_q^{(\tau)}}$. In the thermodynamic limit, $N\rightarrow \infty$, the transformation can be iterated an arbitrary number of times to obtain a sequence of Hamiltonian matrices
\begin{equation}
\left( {\mathcal H^{(0)} } \right) \stackrel{(u^{(1)} ,w^{(1)} )}{\longrightarrow} \left( {\mathcal H^{(1)} } \right) \stackrel{(u^{(2)} ,w^{(2)} )}{\longrightarrow} \left( {\mathcal H^{(2)} } \right)\stackrel{(u^{(3} ,w^{(3)} )}{\longrightarrow}  \ldots  \label{s6e9}
\end{equation}
each describing a theory with quadratic interactions between at most next-nearest-neighbor sites, and each defined on an identical lattice $\mathcal L$ that has $M$ bosonic modes per lattice site. 

\subsection{Ground state RG} \label{Sec:GroundRG}
In the previous section an implementation of real-space RG that could be used to coarse-grain harmonic lattice Hamiltonians was described. We now detail a similar procedure which allows \emph{ground-states} of the harmonic lattices to be coarse-grained directly. Since we are dealing with systems of free-particles, the covariance matrix $\gamma$ gives a complete description of the (Gaussian) ground state  $\left| {\psi _{{\rm{GS}}} } \right\rangle$. In the case of a Hamiltonian as in Eq. \ref{s6e3}, the convariance matrix is of the form $\gamma  = \gamma _p  \oplus \gamma _q $, where $\gamma _p$  and $\gamma _q$ are defined as
\begin{align}
 \left( {\gamma _p } \right)_{ij}  &\equiv 2\left\langle {\psi _{{\rm{GS}}} } \right|\hat p_i \hat p_j \left| {\psi _{{\rm{GS}}} } \right\rangle  \nonumber\\ 
 \left( {\gamma _q } \right)_{ij}  &\equiv 2\left\langle {\psi _{{\rm{GS}}} } \right|\hat q_i \hat q_j \left| {\psi _{{\rm{GS}}} } \right\rangle.\label{s7e1}   
\end{align}
The derivation of analytic expressions for $( \gamma_p$, $\gamma_q )$ is a standard calculation and can be found e.g. Ref. \cite{audenaert}. Again, we choose the disentanglers $u$ and isometries $w$ to be canonical transformations. We obtain a sequence of increasingly coarse-grained states each defined by covariance matrix $\gamma^{(\tau)}$ 
\begin{equation}
\left( {\gamma ^{(0)} } \right) \stackrel{(u^{(1)} ,w^{(1)} )}{\longrightarrow} \left( {\gamma ^{(1)} } \right)\stackrel{(u^{(2)} ,w^{(2)} )}{\longrightarrow} \left( {\gamma ^{(2)} } \right)\stackrel{(u^{(3)} ,w^{(3)} )}{\longrightarrow} \ldots \label{s7e2}
\end{equation}
with $\gamma^{(0)} \equiv \gamma$ as the original ground state. The coarse-graining transformations of the ground state shall be realized by symplectic transforms $S$ acting in the space of the covariance matrix, $\gamma  \mapsto \gamma ' = S^T \gamma S$. As there is no correlation between $\hat p$ and $\hat q$ quadrature coordinates, i.e. $\gamma  = \gamma _p  \oplus \gamma _q $, we may restrict consideration to symplectic transforms $S\in\textrm{Sp}(2N,\mathbb R)$ that are only of the form $S = A \oplus (A^{ - 1} )^\textrm{T}$ with $A$ a real, invertible $N\times N$ invertible matrix. The covariance matrix $\gamma$ transforms under conjugation by matrix $S$, which implies that
\begin{align}
 \gamma _p  &\mapsto \gamma '_p  = A^T \left( \gamma _p \right) A \nonumber\\ 
 \gamma _q  &\mapsto \gamma '_q  = A^{ - 1} \left( \gamma _q \right) (A^{ - 1} )^T, \label{s7e3}  
\end{align}
yeilding a new state $\gamma ' = \gamma_p ' \oplus \gamma_q '$. The disentanglers $u$, which act on two contiguous sites each of $M$ modes, are thus realized as real, invertible $2M\times 2M$ matrices that transform the covariance matrix as per Eq. \ref{s7e3}. Isometries $w$, which act on two contiguous sites within a block, are realised as $w = w_0 w_\textrm{proj}$ with $w_0$ as a real, invertible $2M\times 2M$ matrix and $w_\textrm{proj} = (0_M  \oplus I_M)$ a projection onto $M$ modes of the block. The direct-sum of the local operators $(u,w)$ is constructed
\begin{equation}
 W_\textrm{proj}  =  {\mathop  \bigoplus \limits_{i = 1}^{N/2L} w_\textrm{proj} } ,\ \ W_0  = {\mathop  \bigoplus \limits_{i = 1}^{N/2L} w_0 }, \label{s7e4}    
\end{equation}
\begin{equation}
 U =  {\mathop  \bigoplus \limits_{i = 1}^{N/2L} u} . \label{s7e5} 
\end{equation}
in order to coarse-grain the entire lattice. Starting from the covariance matrix $\gamma = \gamma_p \oplus \gamma_q$ describing the ground state of the original system and, for any choice of $(u,w)$, a new state $\gamma' = \gamma_p ' \oplus \gamma_q '$ can be obtained on a coarser lattice $\mathcal L '$ through an ER transform
\begin{align}
 \gamma _p '  &= \left( W_{{\rm{proj}}} W_0^\textrm{T} U^\textrm{T} \right) \gamma _p \left( U W_0 W_{{\rm{proj}}} \right),  \nonumber \\
 \gamma _q '  &= \left( W_{{\rm{proj}}} W_0^{ - 1} U^{-1} \right)           \gamma _q \left( (U^{-1})^\textrm{T} (W_0^{ - 1})^\textrm{T} W_{{\rm{proj}}} \right). \label{s7e6}   
\end{align}
It is only for proper choice of tensors $(u,w)$ that the above transformation correctly preserves the properties of the original state and produces a meaningful coarse-grained state. We now address the issue of how the proper tensors $(u,w)$ can be found. In the application of the RG to the system Hamiltonian, both in momentum-space and real-space formulation, the modes truncated at each iteration were chosen as \emph{high-energy modes} in order to leave the low-energy structure of the original theory intact. For the ground state RG a different criteria is required to judge which modes should be truncated from the system. The proper truncation criteria in order to preserve the ground state properties, proposed by White as part of his DMRG algorithm \cite{dmrg}, requires that the truncation of a block should be chosen to keep the support of the density matrix for the block. In the present formulation of bosonic modes, in which the representation of the state is given by a covariance matrix as opposed to a density matrix, this rule imposes that only modes in a block that can be identified as being in a \emph{product state} with the rest of the system can be truncated and safely removed from the description of the state. Thus in comparison with Hamiltonian RG, which was optimised to truncate modes such that effective theory had minimal energy, here we optimise for $(u,w)$ so that the truncated modes have minimal entanglement with the rest of the system.

The entanglement of a block with the rest of the system is known to be related to the symplectic eigenvalues of $\gamma$ for the block \cite{plenio,audenaert}. Let the covariance matrix $(\gamma) |_{\mathcal B} = ({\gamma _p}) |_\mathcal{B} \oplus ({\gamma_q}) |_\mathcal B $ describe the correlations within a block $\mathcal B$ of two $M$-mode sites. The $2M$ symplectic eigenvalues $\lambda_i$ of the block $\mathcal B$ are the eigenvalues of the matrix formed by taking the product of matrices $({\gamma _p}) |_\mathcal{B}$ and $({\gamma_q}) |_\mathcal B$
\begin{equation}
\lambda _{i}  = {\rm{Spect}}\left\{ ({\gamma _p}) |_\mathcal{B} ({\gamma_q}) |_\mathcal B \right\}.\label{s7e7}
\end{equation}
The Heisenberg position-momentum uncertainty relation, which here may be simplified as $\left\langle {\hat p^2 } \right\rangle \left\langle {\hat q^2 } \right\rangle  \ge 1/4$, enforces that all symplectic eigenvalues are positive and have magnitude greater than unity, $\lambda_i\ge 1$ for all $i$. The entanglement entropy $S_i$ of a mode `$i$' with symplectic eigenvalue $\lambda_i$ is
\begin{equation}
S_{i}  = \left[ {f\left( {\frac{{\sqrt {\lambda _i }  - 1}}{2}} \right) - f\left( {\frac{{\sqrt {\lambda _i }  + 1}}{2}} \right)} \right] \label{s7e8}
\end{equation}
with $f(x) =  - x\log x$. It is seen that the entropy $S_i$ is zero when mode $i$ is in a minimum uncertainty state, $\lambda_i=1$, and an also that $S_i$ is \emph{increasing} function of $\lambda_i$. It follows that, if a mode with eigenvalue $\lambda_i=1$ can be identified within a block $\mathcal B$, then we are assured that the mode is in a product state with the rest of the system and may be safely truncated. Hence the tensors $(u,w)$ should be optimized to minimize the eigenvalues $\lambda_i$, and thus the entanglement entropy, of the modes to be projected out. The projection $W_\textrm{proj}$ in Eq. \ref{s7e4} was defined to select the modes to be retained; we now construct the complimentary projector $\tilde W_\textrm{proj}$
\begin{equation}
\tilde W_\textrm{proj}  =  {\mathop  \bigoplus \limits_{i = 1}^{N/2M} \left( {I_{2M}  - w_\textrm{proj} } \right)} \label{s7e9}
\end{equation}
which projects onto the space of the modes to be truncated. The part $\tilde \gamma$ of the covariance matrix $\gamma$ that is to be projected out during the RG iteration may be written $\tilde \gamma = \tilde \gamma_p \oplus \tilde \gamma_q$ with
\begin{align}
 \tilde \gamma _p  &= \left( \tilde W_{\textrm{proj}} W_0^\textrm{T} U^\textrm{T} \right) \gamma _p \left( U W_0 \tilde W_{\textrm{proj}} \right) \nonumber \\ 
 \tilde \gamma _q  &= \left( \tilde W_{{\rm{proj}}}  {W_0^{ - 1} }   U^{-1} \right)       \gamma _q \left( (U^{-1})^\textrm{T} (W_0^{ - 1})^\textrm{T} \tilde W_{{\rm{proj}}}\right). \label{s7e10}
\end{align}
Given that the modes projected out of block $\mathcal B$ are to have minimum entropy, the transforms $(u,w)$ should be chosen to minimise the matrix equation
\begin{equation}
\mathop {\min }\limits_{u,w} \left( {\textrm{tr}}\left\{ {(\tilde \gamma _q ) }|_\mathcal{B}  {(\tilde \gamma _p )  }| _\mathcal{B}  \right\} \right). \label{s7e11}
\end{equation}
As with the energy minimization described by Eq. \ref{s6e8} for the Hamiltonian RG , this equation is optimised variationally to find good disentanglers $u$ and isometries $w$.

\begin{figure}[!tb]
  \begin{center}
    \includegraphics[width=6cm]{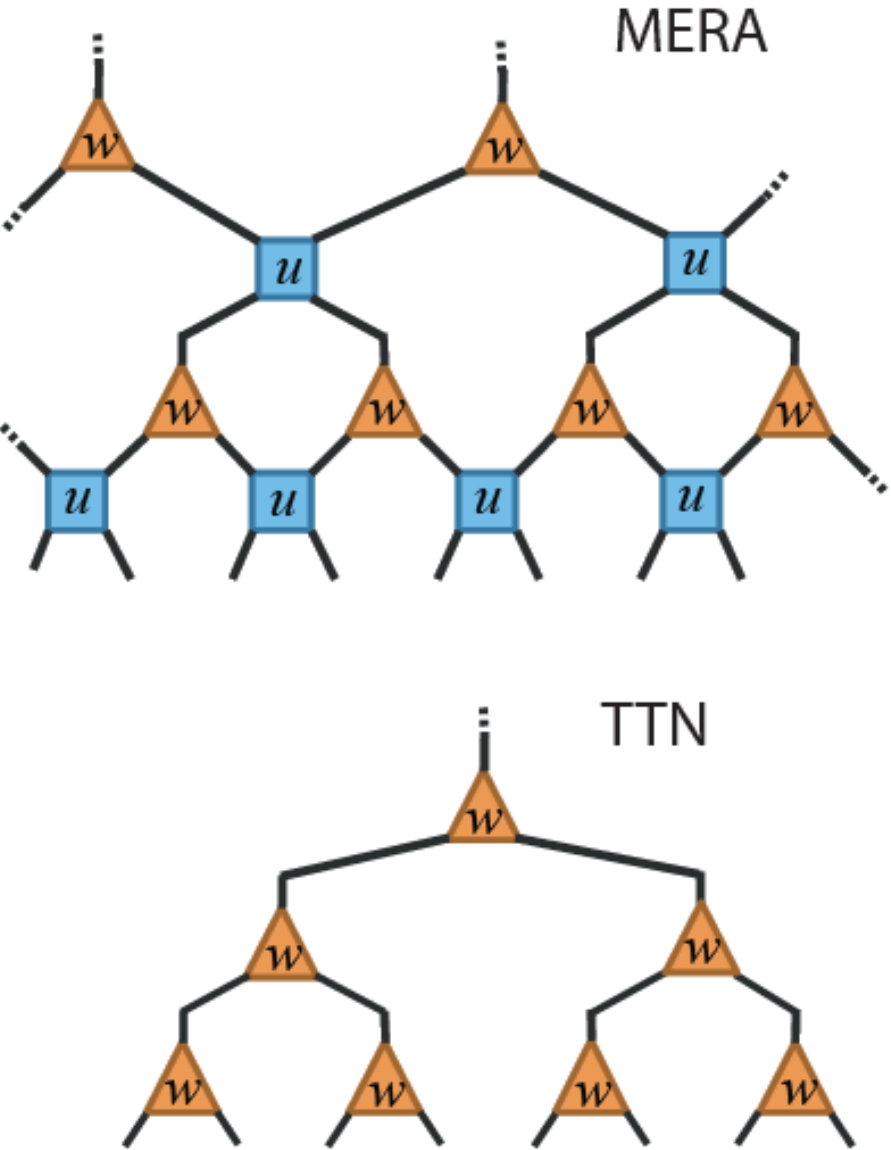}
  \caption{(Top) Applying successive RG maps based upon ER, as depicted in Fig. \ref{ProjMera}, to the ground state leads to an approximate representation of the ground state in terms of a set of disentanglers and isometries $(u^{(\tau)}, w^{(\tau)})$ connected in a tensor network. This tensor network forms the \emph{multi-scale entanglement renormalization ansatz} (MERA) \cite{MERA}, an ansatz for many-body states on a lattice. (bottom) Implementing the RG based upon an LP coarse-graining gives an approximation to the state in terms of a \emph{tree-tensor network} (TTN) \cite{TTN1, TTN2}, a different class of ansatz for many-body states on a lattice.}  
  \label{MERAvsTTN}
 \end{center}
\end{figure}

The application of RG transformations to the Hamiltonian could be interpreted as producing effective theories for the low-energy subspace of the original system. Is there a similar interpretation for the coarse-graining of the ground state? To address this question we assume that the ER map has been iterated $\tau$ successive times on the original ground state $\gamma^{(0)}$ to get the coarse-grained state $\gamma^{(\tau)}$. Each of the $\tau$ coarse-grainings is characterised by a disentangler $u$ and isometry $w$, thus the set of these transforms $\left( u^{(j)}, w^{(j)} \right)$ for $j = 1,2,\ldots,\tau$ characterise the sequence of RG maps. If the modes truncated during each iteration were in an exact product state, equivalently the eigenvalues of every mode `$i$' truncated was $\lambda_i = 1$, then the sequence of transformations could be inverted as follows. Starting from $\gamma^{(\tau)}$, truncated modes (which were in a product state) are replaced back, and each of the transforms of Eq. \ref{s7e6} is inverted
\begin{equation}
\left( {\gamma ^{(\tau)} } \right) \stackrel{(u^{(\tau)} ,w^{(\tau)} )}{\longrightarrow} \left( {\gamma ^{(\tau-1)} } \right)\stackrel{(u^{(\tau-1)} ,w^{(\tau-1)} )}{\longrightarrow} \ldots \stackrel{(u^{(1)} ,w^{(1)} )}{\longrightarrow} \left( {\gamma ^{(0)} } \right), \label{s7e12}
\end{equation}
as to recover the exact original state $\gamma^{(0)}$. The coarse-graining of the ground state can be interpreted as \emph{storing} information about the short range properties of the state into the tensors $\left( u^{(i)}, w^{(i)} \right)$ while preserving the long range information about the state; the set of tensors $\left( u^{(i)}, w^{(i)} \right)$ together with state $\gamma^{(\tau)}$ thus serve as a representation of the original state $\gamma^{(0)}$. The set $\gamma^{(\tau)}$ and $\left( u^{(i)}, w^{(i)} \right)$ form the \emph{multi-scale entanglement renormalization ansatz} (MERA) \cite{MERA}, a variational ansatz for states on the lattice, c.f. Fig \ref{MERAvsTTN}. [If the ground state RG is performed with an LP coarse-graining, as opposed to one based upon ER, a \emph{tree tensor network} (TTN) \cite{TTN1, TTN2} approximation to the state is obtained in terms of the isometries $ w^{(i)} $]. For the harmonic lattices we consider, as with most non-trivial models, the coarse-graining cannot be performed exactly and a MERA will be an approximate, rather than exact, representation of the true ground state. In the present case, irrespective of how the transforms $\left( u, w\right)$ are chosen, the modes to be truncated will still be slightly entangled as manifested in their symplectic eigenvalues, which will fulfill $\lambda_i > 1$. This entanglement will be ignored during the coarse-graining thus limiting the precision with which the original ground state $\gamma^{(0)}$ may be recovered.

In the present setting of free bosons, in which the exact ground state $\gamma^{(0)}$ is already known, we compute the MERA and TTN approximations to the ground state via coarse-graining the ground state covariance matrix. 

The point of this exercise is that it allows us to assess the accuracy with which a MERA or a TTN can represent the many-body state. It also provides the opportunity to study the RG flow of the ground state $\gamma^{(\tau)}$ under scale transformations. However, had our goal been to investigate properties of the unknown ground state of a system $\hat H$ with the help of real-space RG, it would have been absurd to assume knowledge of the exact ground state $\left| {\psi _{{\rm{GS}}} } \right\rangle$ from the beginning. In that case, an approximation $\left| {\tilde \psi _{{\rm{GS}}} } \right\rangle$ to the ground state may be found through e.g. variational minimization of energy $\left\langle {\tilde \psi _{GS} } \right|\hat H\left| {\tilde \psi _{GS} } \right\rangle$ as per Ref. \cite{algorithms} or through an alternative method \cite{unified, tMERA}.

\begin{figure}[tb]
  \begin{center}
    \includegraphics[width=8cm]{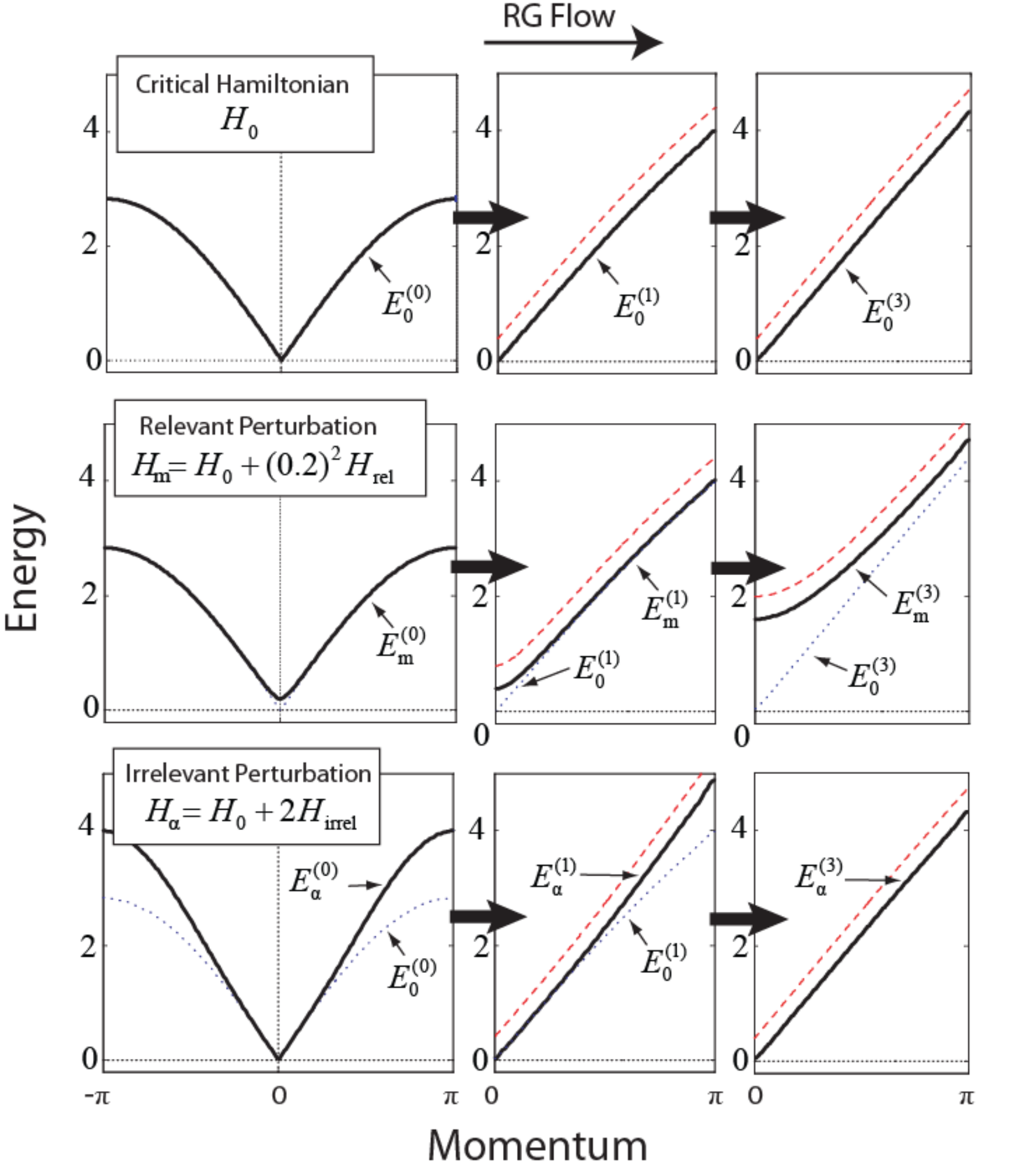}
  \caption{Sequences of dispersion relations of Hamiltonians and after $\tau=0,1,3$ RG transforms, comparing the real-space ER results (bold) with the exact momentum-space results (dashed, offset by 0.2). (Top Series) The critical Hamiltonian $\hat H_0$ tends to a linear dispersion under the RG map with ER, in good agreement with momentum-space results. (Middle Series) The addition of even small mass, $m=0.2$, to the critical system gives a marked difference in the dispersion relation after $\tau=3\ $ RG transforms. (Bottom Series) The effect of an irrelevant perturbation quickly diminishes under RG flow; by $\tau=3$ iterations the original and the perturbed dispersions are virtually identical, $E_{{\alpha}}^{(3)} \approx E_0^{(3)}$.}
  \label{BoseHamRenorm}
 \end{center}
\end{figure}

\begin{figure}[tb]
  \begin{center}
    \includegraphics[width=8cm]{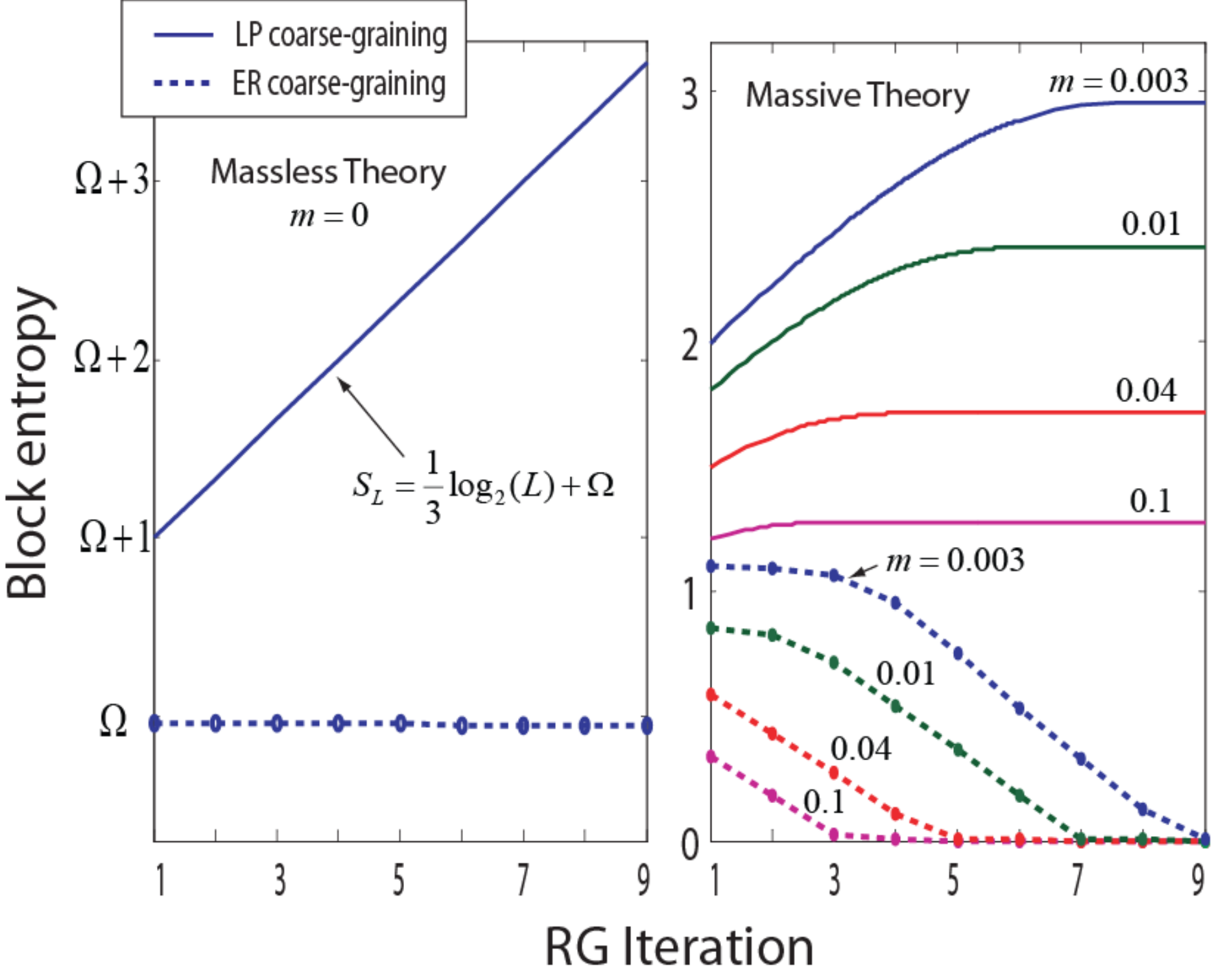}
  \caption{The entanglement entropy $S$ of a site in the $\tau^\textrm{th}$ coarse-grained ground state $\gamma^{(\tau)}$ of the $1D$ harmonic chain. (Left) In the critical (massless) regime the entropy of a site is infinite, as realized by the infinite constant $\Omega$; however the change in entropy along the RG flow can be computed via a limiting process. The entropy of the state renormalized with an LP coarse-graining increases by a constant with each RG iteration, reproducing the logarithmic growth law, $S_L = (1/3)\log_2 L + c$, as expected from conformal field theory \cite{logVidal, Cala}, whilst the entropy of the system renormalized with ER remains constant along the RG flow. Furthermore the sequence of renormalized ground-states $\{ {\gamma ^{(1)} ,\gamma ^{(2)} , \ldots } \}$ rapidly converge to a fixed state $\gamma^*$ under the ER transforms. (Right) For several values of finite mass, the entropy of the states renormalized with the LP scheme saturate at a length scale governed by correlation length, whilst the theories renormalized with ER factorize into a product state (zero entropy) at the approximately same length. }
  \label{BoseEntPlotFlat}
 \end{center}
\end{figure}

\begin{figure}[!tb]
  \begin{center}
    \includegraphics[width=8cm]{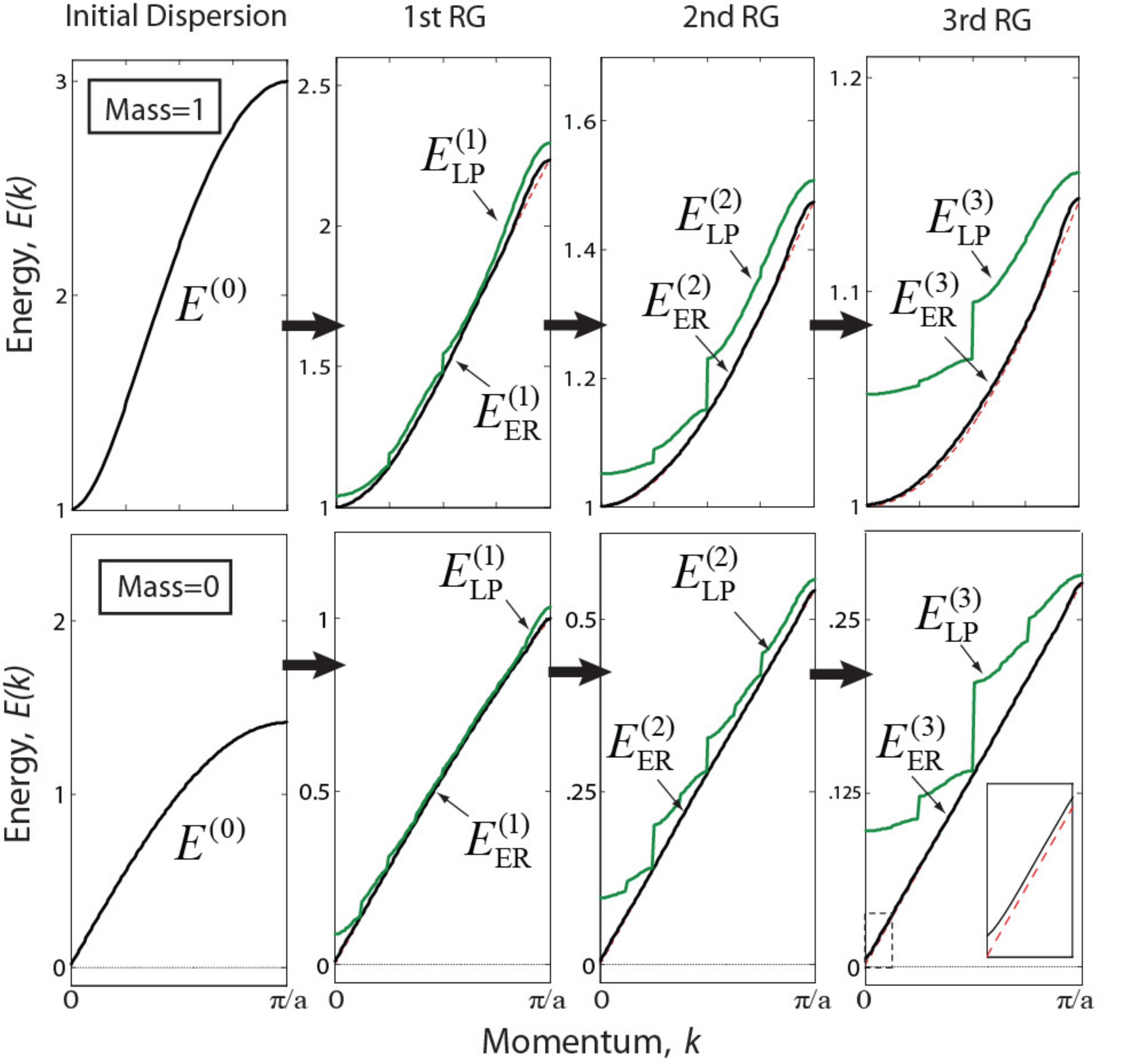}
  \caption{(Left) Sequences of dispersion relations (in non-rescaled energy units) for the (top) gapped and (bottom) critical $1D$ harmonic systems after $\tau=0,1,2,3$ real-space RG transforms, comparing the results from the local projection (LP) method, $E_\textrm{LP}^{(\tau)}$, to those from the entanglement renormalization (ER) coarse-graining, $E_\textrm{ER}^{(\tau)}$. The numeric dispersions produced by ER agree with the exact results obtained from momentum-space RG (dashed) as to be almost visually indistinguishable; though small errors are noticeable near the momentum cut-off, $k=\pi/a$. The dispersion relations given by coarse-graining performed with the LP scheme, whilst reasonably accurate after the $1^\textrm{st}$ iteration, rapidly diverge from the exact result with successive RG iterations. In all cases the numeric RG was performed keeping $M=4$ modes per site. }
  \label{BoseHamError}
 \end{center}
\end{figure}

\begin{figure}[!tb]
  \begin{center}
    \includegraphics[width=8cm]{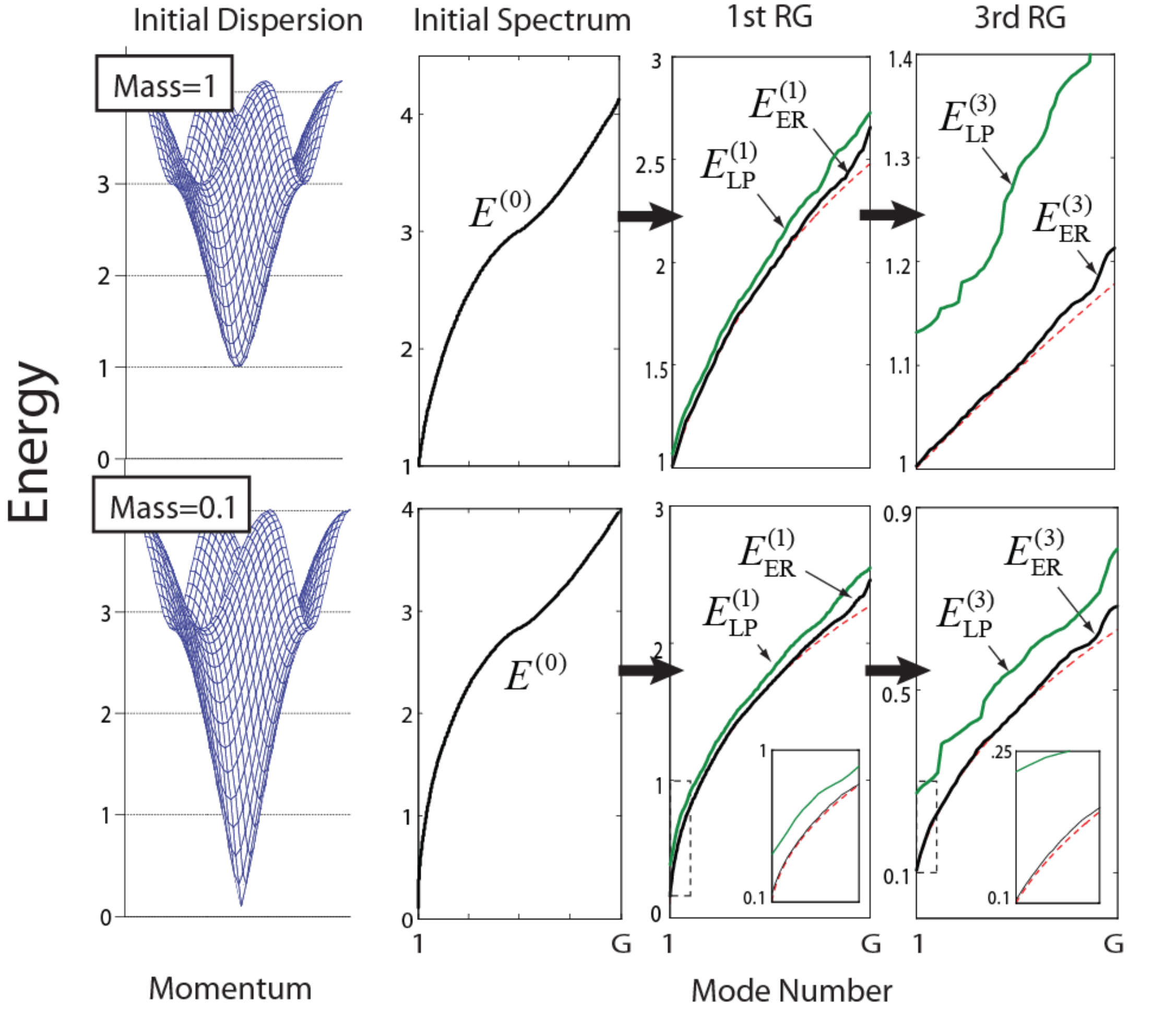}
  \caption{Sequences of energy spectra (in non-rescaled energy units) for the (top) gapped and (bottom) near-critical $2D$ harmonic lattice system after $\tau=0,1,3$ real-space RG transforms. The performance of the numeric real-space methods, local projection (LP) and entanglement renormalization (ER), are benchmarked against the exact solutions from momentum-space RG (dashed). The energy spectra are obtained by sampling the dispersion relation $E(k_1,k_2)$ on a finite grid of $G$ points, with $G$ chosen very large, and then ordering the values, $\{E_1\le E_2\le \cdots \le E_G\}$. The spectra of the systems renormalized with entanglement renormalization, $E_\textrm{ER}^{(\tau)}$, retain a high level of accuracy through all $\tau=3$ RG iterations, though do lose precision near the high energy cut-off of the effective theory. Energy spectra $E_\textrm{LP}^{(\tau)}$ of the effective theories obtained with an LP coarse-graining have diverged considerably from the exact result by $\tau=3$ RG iterations. The numeric RG was performed keeping $M=9$ modes per site.  }
  \label{BoseError2D}
 \end{center}
\end{figure}

\section{Results and Discussion} \label{Sec:Results}
\subsection{RG flow of Hamiltonians} \label{Sec:HamResult}

In Sect. \ref{Sec:MRG} the low energy subspace of harmonic chains were analysed exactly with momentum-space RG. This analysis yielded sequences of dispersion relations describing the RG flow towards a critical fixed point as well as the RG flow resulting from adding relevant and irrelevant perturbation terms, as given by Eqs. \ref{s2e7}, \ref{s3e3} and \ref{s4e3} respectively. The same harmonic systems can also be analysed numerically by applying successive real-space RG transformations $\mathcal{T}$, based upon ER, following the variational method described in Sect. \ref{Sec:HamRG}. This produces a sequence of Hamiltonian matrices $\left( \mathcal{H}^{(0)}, \mathcal{H}^{(1)}, \mathcal{H}^{(2)}, \ldots \right)$, each an effective theory describing successively lower energy subspace of the original system. By construction, $\mathcal{H}^{(\tau)}$ only contains quadratic couplings between nearest and next-to-nearest neighboring sites. The dispersion relations of the effective Hamiltonians $\mathcal H^{(\tau)}$ are found by Fourier transform in a similar manner as presented for the original Hamiltonian in Sect. \ref{Sec:Harm}. Figure \ref{BoseHamRenorm} shows the comparison of the analytic (momentum-space) and numeric (real-space) dispersion relations.

The dispersion relations produced from numeric real-space RG with ER are seen to approximate the exact results to a high degree of accuracy for the three of the cases considered: a critical system, and relevant and irrelevant perturbations on the critical system. Furthermore, the critical Hamiltonian $\mathcal H_0^{(0)}$ converges to a fixed point of the RG flow, $\mathcal{H}_0^{(\tau)} = \mathcal{H}^{*}$, after $\tau \ge 3$ RG transforms. This fixed point Hamiltonian matrix $\mathcal{H}^{*}$ is manifestly invariant under further transformations, $\mathcal{T} (\mathcal {H}^{*}) = \mathcal {H}^{*}$, to within small numerical errors. Consequently the isometries $w$ and disentanglers $u$, which comprise the ER transform $\mathcal{T}$, also converged to fixed points $w^{(\tau)} = w^*$ and $u^{(\tau)} = u^*$ for $\tau \ge 3$. The Hamiltonian matrix $\mathcal{H}_\alpha^{(0)}$ of the critical system with the addition of an irrelevant perturbation, $\hat H_\alpha^{(0)} = \hat H_0^{(0)} + \alpha \hat H_\textrm{irrel}^{(0)}$, converges to the same fixed point as the unperturbed Hamiltonian $\mathcal{H}_\alpha^{(\tau)} = \mathcal{H}_0^{(\tau)} = \mathcal{H}^{*}$ for $\tau \ge 3$ transforms.

\subsection{RG flow of ground states} \label{Sec:GroundResult}

On the other hand, in Sect. \ref{Sec:GroundRG} we describe a real-space RG to coarse-grain the ground state of the $1D$ harmonic chain. The coarse-graining transformations can be based upon either the LP or ER schemes of Fig. \ref{ProjMera}. The real-space RG produces a sequence of increasingly coarse-grained ground-states $\left( \gamma^{(0)}, \gamma^{(1)}, \gamma^{(2)}, \ldots \right)$, where $\gamma^{(0)}$ is the ground state covariance matrix of a harmonic chain with mass $m$. Fig. \ref{BoseEntPlotFlat} displays the entanglement entropy $S$, as defined in Eq. \ref{s7e8}, of a site in the coarse-grained state $\gamma^{(\tau)}$ as a function of the RG iteration $\tau$.

The entropy of the ground state $\gamma_0^{(\tau)}$ of the critical chain, $m=0$, when coarse-grained with the LP scheme, increases by a constant with each RG iteration. More precisely, if we recall that a site in lattice $\mathcal L ^{(\tau)}$ corresponds to a block of $L = 2^\tau$ sites of the original lattice, this growth of entropy reproduces the expected logarithmic scaling, $S_L = (1/3) \log_2 L + c$, for $1D$ critical systems \cite{logVidal, Cala}. This growth demonstrates that it is impossible for a critical ground state to be a fixed point of the LP scheme.

The ground state of a system with finite mass $m$ can also be analysed with the LP method. Initially, the ground state displays the same logarithmic growth of entropy as in the critical case, but it saturates approximately after $\tau = \log_2 \zeta$ iterations of the RG transformation, where $\zeta$ is the correlation length.

Turning to the ER scheme, if disentanglers are included in the coarse-graining step then the entropy per site of state $\gamma_0^{(\tau)}$ remains constant under RG transforms. This is made possible by the disentanglers, which remove short-range entanglement at each iteration. Moreover, for the critical case, the sequence of successively coarse-grained ground-states $\gamma_0^{(\tau)}$ explicitly converges to a fixed point, $\gamma_0^{(\tau)}=\gamma^*$ for $\tau \ge 3$. The non-critical ground states $\gamma^{(\tau)}$ of a harmonic chain with finite mass $m$ converged to a trivial fixed point, a product state $\gamma^{**}$, under RG iteration. This occurs at the length scale of the correlation length $\zeta$. The numeric results have been obtained by keeping $M=4$ bosonic modes per lattice site. The MERA approximation to the ground state, obtained from successive coarse-grainings with ER, proves remarkably accurate. For the critical system, which is the hardest to analyse from a computational point of view, the MERA obtained from $\tau =9$ RG iterations can reproduce the exact correlators of the ground state to error bounded by $1\times 10^{-4}$.

\begin{figure}[!tb]
  \begin{center}
    \includegraphics[width=8cm]{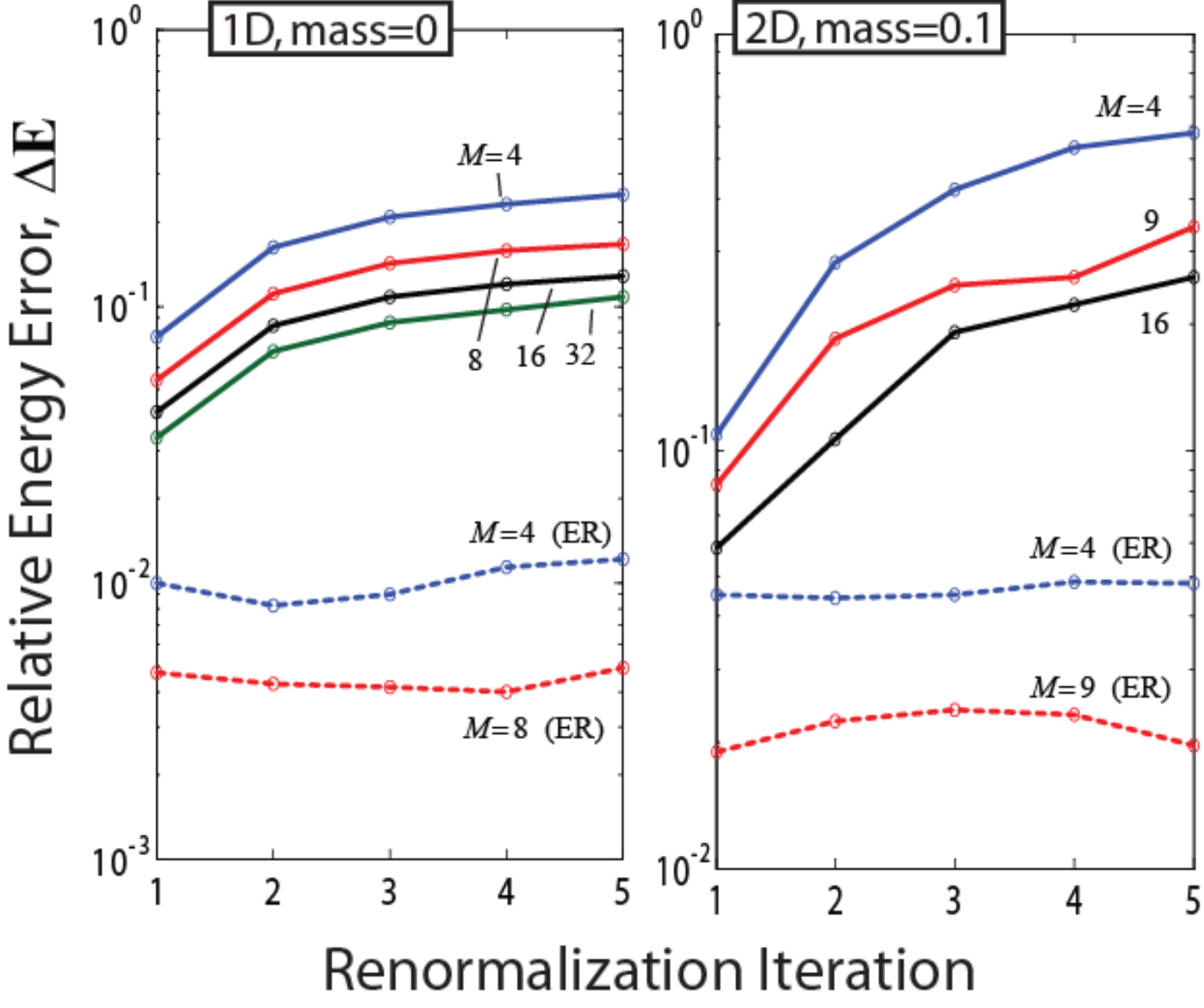}
  \caption{The mean energy $\bar{E}_\textrm{RS}$, as defined Eq. \ref{s8e1}, of the effective theories obtained through real-space RG, based upon either an LP (solid) or ER (dashed) coarse-graining, are compared against exact momentum-space results $\bar{E}_\textrm{MS}$ in terms of the relative error $\Delta {E} = (\bar{E}_\textrm{RS} - \bar{E}_\textrm{MS}) / \bar{E}_\textrm{MS}$. Performing the real-space coarse-graining with a larger number of modes per site $M$, which allows for more parameters to be kept in the description of the effective theories, is shown to give more accurate results for the real-space RG. (left) For the $1D$ critical harmonic chain the mean of the energy spectrum obtained from renormalizing with ER, while keeping $M=4$ modes per site, remains approximately $1\%$ greater than exact value throughout the RG iterations. The spectra obtained from the LP method, even when using significantly larger $M$, only gave at best $10\%$ accuracy after the same number of RG iterations. (right) For the $2D$ near-critical harmonic lattice, ER gives accuracies after 5 iterations of no less than $4\%$ and $2\%$ for $M=4$ and $M=9$ respectively; this is in contrast to the LP method which is only accurate to within $60\%$ for $M=4$ and $30\%$ for $M=9$. Barring small fluctuations, the accuracy of the coarse-graining with ER remains relatively stable with RG iteration. }
  \label{1D2Derrraw}
 \end{center}
\end{figure}

\subsection{Local Projection vs Entanglement Renormalization} \label{Sec:LPvsER}

Figs. \ref{BoseHamError} and \ref{BoseError2D} present numeric dispersion relations for the harmonic lattices in $D=1$ and $D=2$ spatial dimensions respectively, and compare results produced by coarse-graining with LP to those obtained by coarse-graining with ER. The dispersion relations for the effective theories produced from the LP method are reasonably accurate after a small number of RG iterations. However, they rapidly diverge from the exact results in subsequent iterations. On the other hand, coarse-graining with ER is shown to keep significantly better support of the low energy subspace and, most importantly, to maintain accuracy over repeated RG iterations. The numeric spectra obtained with ER for the $2D$ lattices of Fig. \ref{BoseError2D} displays a loss of accuracy towards the high-energy cut-off, which indicates difficulty in keeping a sharp cut-off numerically. However, this is of little concern as the primary interest lies in the low-energy physics and not the high-energy cut-off of the effective theory.

The average mode energy $\bar E$ of an effective theory, defined in terms of its dispersion relation $E(k)$, is
\begin{equation}
\bar{E} \equiv \frac{1}{{2\Lambda }}\int_{ - \Lambda }^\Lambda  {E(k)dk} \label{s8e1}
\end{equation}
for a $1D$ system and similar for $2D$. The average mode energy is used to qualitatively analyze the accuracy of the effective theories obtained with real-space RG, as presented in Fig. \ref{1D2Derrraw}. Keeping more modes-per-site $M$, hence more parameters in the description of the effective theory, increases the accuracy with which the numeric RG may be performed. However, even with the choice of a very large $M$, very large the LP method shows significant increase in error along the RG flow for both the $1D$ critical and $2D$ near-critical harmonic systems. This figure also confirms what was established visually in Figs. \ref{BoseHamError} and \ref{BoseError2D}; that coarse-graining with ER not only produces a more accurate low-energy theory than with LP, but also maintains constant accuracy over successive RG maps.  

\section{Conclusions} \label{Sec:Conclude}
Real-space RG techniques, primarily in the form of the DMRG algorithm \cite{dmrg, DMRGreview}, have proved an invaluable tool for the numeric analysis of low-energy properties of extended quantum systems. However, this class of real-space RG methods (including the LP coarse-graining described here) suffer from an important deficiency: namely they do not reproduce proper RG flows. As demonstrated by Fig. \ref{BoseEntPlotFlat}, a $1D$ critical system could not possibly be a fixed point of the LP map due to the growth of entropy along the RG flow, which is caused by the accumulation of short-range degrees of freedom. This deficiency in turn limits the size of $1D$ critical and $2D$ lattices that can be analysed with such methods.      

In this work we have demonstrated, by comparison against exact results from momentum-space RG, that a coarse-graining based upon entanglement renormalization can reproduce proper RG flows. Demonstrations included the analysis of a critical Hamiltonian $\mathcal H_0$ and its ground state $\gamma_0$, which were shown to rapidly converge to non-trivial fixed points of the RG flow, $\mathcal H_0^*$ and $\gamma_0^*$ respectively, and analysis of several non-critical systems which converged to trivial fixed-points of the RG flow. By addition of an irrelevant perturbation to the critical Hamiltonian $\mathcal H_0$, a new Hamiltonian $\mathcal H_\alpha$ was obtained that described the same phase as $\mathcal H_0$, but differed in the local interaction. As is required of a proper RG flow, $\mathcal H_\alpha$ and $\mathcal H_0$ converged to the same fixed point $\mathcal H_0^*$ under the RG map defined by ER. It is important to note that the tensors $(u,w)$, which comprised the real-space RG transform, were not chosen based on heuristic arguments or the desire for a particular outcome. Instead they were found through optimisation based upon either energy minimization, as in Eq. \ref{s6e8}, or attempting to retain the support of the ground state, as in Eq. \ref{s7e11}.

In addition, the coarse-graining transformation based upon ER was shown to induce a \emph{sustainable} RG map; one that could be applied arbitrarily many times without significant loss of accuracy and without the need to increase the local dimension $M$ of the effective theories (so that the computational cost is also kept constant). The sustainability of the ER based RG map allows investigation of low-energy properties in arbitrarily large or infinite $1D$ and $2D$ lattices, as has also been demonstrated in recent studies in which local observables are evaluated directly in the thermodynamic limit \cite{algorithms, ER2D, ERKag} and critical exponents computed \cite{CFT,CFT2} without the need for finite size scaling techniques. 

The implementation of entanglement renormalization presented in this work exploited properties of free-particle systems in order to reduce the computational cost. More general algorithms exist \cite{unified, tMERA, algorithms} which allow implementation without making use of the special properties of free particle systems. Thus it is possible to use real-space RG, based upon ER, as a means of investigating the low-energy properties of strongly-correlated systems where perturbative approaches are not valid. However, since the implementation of ER in the interacting case is computationally expensive, especially so for $2D$ lattices, it remains unclear the level of accuracy that may be obtained in practice. Early investigations of $2D$ lattice systems with the full algorithm are promising \cite{cincio,ER2D,ERKag,Ferm2,Ferm3,Ferm4}, and the development and application of entanglement renormalization remains a subject of ongoing research.

Financial support of the Australian Research Council (APA, FF0668731, DP0878830) is acknowledged.\\

\end{document}